\PassOptionsToPackage{svgnames}{xcolor}
\documentclass[%
nofootinbib,
nobibnotes,
12pt
]{iopart}

\usepackage{hhline}

\usepackage{tikz}
\usetikzlibrary{plotmarks}

\usepackage{xcolor}

\usepackage{graphicx,caption,subcaption}
\usetikzlibrary{decorations.pathmorphing}

\usepackage{dcolumn}
\usepackage{bm}

\usepackage{hyperref}

\usepackage{mwe}

\usepackage{caption}
\usepackage{subcaption}

\renewcommand{\thesection}{}
\renewcommand{\thesubsection}{\arabic{section}.\arabic{subsection}}
\makeatletter
\def\@seccntformat#1{\csname #1ignore\expandafter\endcsname\csname the#1\endcsname\quad}
\let\sectionignore\@gobbletwo
\let\latex@numberline\numberline
\def\numberline#1{\if\relax#1\relax\else\latex@numberline{#1}\fi}
\makeatother

\usepackage{blindtext}
\usepackage [english]{babel}
\usepackage [autostyle, english = american]{csquotes}
\MakeOuterQuote{"}

\usepackage{titlesec}
\titleformat{\section}
  {\normalfont\sffamily\large\bfseries\color{black}}
  {\thesection}{1em}{}

\hypersetup{colorlinks=true,urlcolor=blue}

\renewcommand{\thesubsection}{\Alph{subsection}}

\usepackage{etoolbox}
\makeatletter
\def\@mkboth#1#2{}
\newlength\appendixwidth
\preto\appendix{\addtocontents{toc}{\protect\patchl@section}}
\newcommand{\patchl@section}{%
  \settowidth{\appendixwidth}{\textbf{Appendix }}%
  \addtolength{\appendixwidth}{1.5em}%
  \patchcmd{\l@section}{1.5em}{\appendixwidth}{}{\ddt}%
}
\makeatother

\usepackage{iopams}

\expandafter\let\csname equation*\endcsname\relax

\expandafter\let\csname endequation*\endcsname\relax

\usepackage{amsmath}

\usepackage{dcolumn}
\usepackage{epsf}
\usepackage{float}

\usepackage{tabularx}

\usepackage[final]{changes}

\usepackage{lipsum}
\definechangesauthor[color=blue]{.}
\colorlet{Changes@Color}{red}
\setremarkmarkup{(#2)}

\begin{document}

\title[]{Exotic quantum statistics \added{and thermodynamics} from \added{a number-conserving} theory of Majorana fermions}

\author{Joshuah T. Heath \& Kevin S. Bedell}

\address{Physics Department,  Boston  College,  Chestnut  Hill, Massachusetts  02467,  USA}
\ead{heathjo@bc.edu}
\vspace{10pt}
\begin{indented}
\item[]\today
\end{indented}

\begin{abstract} \added{We propose a closed form for the statistical distribution of non-interacting Majorana fermions at low temperature.} Majorana particles often appear in the contemporary many-body literature in \added{the Kitaev, Fu-Kane, or Sachdev-Ye-Kitaev models, where the Majorana condition of self-conjugacy immediately results in non-conserved particle number, non-trivial braiding statistics, and the absence of a non-interacting limit.} We deviate from this description and instead consider a gas of non-interacting, spin$-1/2$ Majorana fermions \added{that obey the spin-statistics theorem via imposing a condensed matter analog of momentum conservation.}
%
This allows us to build a quantum statistical theory of the Majorana system in the low temperature, low density limit without the need to account for strong fluctuations in the particle number. A combinatorial analysis leads to a configurational entropy which deviates from the fermionic result with an increasing number of available microstates. A \added{number-conserving} Majorana distribution function is derived which shows signatures of a sharply-defined Fermi surface at finite temperatures. \added{Such a} distribution is then re-derived from a microscopic model in the form of a modified Kitaev chain with a bosonic pair interaction. The thermodynamics of \added{this} free Majorana system is found to be nearly identical to that of a free Fermi gas, except now distinguished by a two-fold ground state degeneracy and, subsequently, a residual entropy at zero temperature. \added{Despite clear differences with the anyonic or Sachdev-Ye-Kitaev models, we nevertheless find surprising agreement between our theory and experimental signatures of Majorana excitations in several materials.} Experimental realization of \added{our exactly solvable model is also discussed in the realm of astrophysical and high-energy phenomena.}

\end{abstract}

\vspace{2pc}

\submitto{\JPA}

\maketitle

\tableofcontents

%


\section{I. Introduction}

\subsection{Background and history}

Dirac's relativistic approach to quantum mechanics, despite correctly predicting spin-orbit coupling and the fine structure of hydrogen \cite{Dirac1, Darwin}, initially faced opposition due to his apparently unphysical "Dirac sea" interpretation of fermionic negative energy states \cite{Dirac2}. Under the encouragement of C.G. Darwin, Eddington was the first to propose an inherently symmetric theory of the Dirac wave equation in the tensor calculus formalism native to special relativity \cite{Eddington, Kilmister}. The symmetric theory of the electron was expanded upon by Ettore Majorana, who re-derived a real variant of the Dirac equation by applying a variational technique to a real field of anti-commuting variables  \cite{Majorana}. In modern notation, the Eddington-Majorana equation is identical to the Dirac equation, except now the complex-valued Dirac matrices generating the $C\ell_{1,\,3}(\mathbb{R})$ Clifford algebra are replaced with purely-imaginary Majorana matrices \cite{Park}. It was Majorana's insight to interpret the solutions to this symmetrized Dirac equation as massive spin-$1/2$ particles identical to their own antiparticle. 

With the detection of the positron providing experimental evidence of a distinct antiparticle state \cite{Anderson}, Majorana's symmetric theory of fermions found popularity in the field of neutrino particles. Essential to Majorana's original theory is that the particles in question are neutral; i.e., that the Eddington-Majorana equation is invariant under charge conjugation \cite{Sivaguru}. As a consequence, Majorana originally proposed the neutron and the neutrino as the most viable realizations of his theory. The former was soon ruled out with the discovery of the antineutron in charge-exchange collisions \cite{Cork}. As for the latter, while it might be possible to detect the emission of an antineutrino in $\beta$ decay, the extremely small neutrino-absorption cross-section of radioactive nuclei renders direct evidence of a Majorana neutrino unlikely \cite{Furry1}. Be that as it may, if the process of double-$\beta$ decay remains absent of neutrino emission, the increased probability of disintegration would be \added{a good} indication that the neutrino is a Majorana fermion \cite{Furry2, Valle}. Although contemporary experiments have yet to detect any signatures of a neutrinoless double-$\beta$ decay \cite{Agostini, GERDA}, experiments at the turn of the century have confirmed the existence of neutrino flavor oscillations and, subsequently, the existence of a non-zero (albeit small) neutrino mass \cite{Fukuda, Ahmad1, Ahmad2}. Such a small mass could be explained via the seesaw mechanism, which assumes a Majorana mass term for the right-handed neutrino on the order of the GUT scale \cite{Mohapatra,Tsutomu,Lavoura}. 

Beyond fundamental particle physics, the idea of a Majorana quasiparticle in a quantum many-body system has become a subject of great interest in the condensed matter community, particularly in the field of superconducting systems \cite{Wilczek, Beenakker3, Wilczek_book, Leggett}. The motivation lies in the form of the Nambu spinor describing a Bogoliubov-de Gennes system with superconducting order, which satisfies the Majorana charge conjugation condition \cite{Elliott}. At zero energy, Majorana quasiparticles form a class of topologically-protected particles known as Majorana zero modes (MZMs) \cite{Flensberg}. MZMs were once thought to only exist in pairs \cite{Volovik, Read} until Kitaev proved in 2001 that a 1D tight-binding chain of spinless fermions in the vicinity of a p-wave superconductor might harbor unpaired MZMs on the chain's boundaries \cite{Kitaev}. Several years later, Fu and Kane showed that edge MZMs can exist as magnetic vortices at the interface of an s-wave superconductor and a strong topological insulator \cite{FuKane}. The topological nature of both the Kitaev and Fu-Kane Majorana quasiparticles have led to the possibility of fault-tolerant quantum computation with MZMs \cite{Bravyi, SuPeng, DasSarma, Pachos}, and has driven researchers to the experimental realization of the former in ferromagnetic atomic chains on the surface of a superconducting lead \cite{Yazdani} and, most recently, a chiral version of the latter in a quantum anomalous Hall insulator--superconductor heterostructure \cite{He}. 

Despite the immense amount of focus on the Majorana zero mode, their physics differs greatly from that of the traditional Majorana fermion \added{studied in high-energy physics}. Kitaev's zero modes are two unlocalized halves of a real fermion that have been confined to the ends of a quantum wire \cite{Wen}, while the Fu-Kane modes associated with point-like topological defects obey the non-Abelian statistics of Pfaffian quantum Hall states \cite{Ivanov, Teo, Zheng}. Even if we were to obtain a hypothetical gas of these Majorana zero modes in some Kitaev model with extended hopping and pairing  \cite{Alecce}, the effects of non-Abelian statistics would become inevitable \cite{Rao}. Majorana zero-energy modes 
can then be thought of as a defining characteristic of topological matter \cite{Read, Cobanera}, whereas Majorana fermions are a natural extension of the particle-hole symmetry and screened Coulomb interactions in a superconducting phase with nonconserved spin \cite{Beenakker1, Senthil}. Consequently, the mutual annihilation of Bogoliubov particles in chiral quantum Hall edge states might be considered a condensed-matter analogy to the neutrinoless double-$\beta$ decay discussed earlier \cite{Beenakker2}. It has even been shown that the electron field amplitudes of planar Dirac-type systems describing s-wave-induced topological superconductivity are described by a Majorana-Eddington wave equation \cite{Chamon}. Beyond superconducting systems, a neutral Majorana Fermi sea has also been suggested to form in the Kondo insulator samarium hexaboride (SmB$_6$) in order to explain unconventional thermodynamic signatures, such as quantum oscillations at low temperatures and a residual thermodynamic entropy \cite{Coleman, Tan, Baskaran}. Nevertheless, there is yet to be a theory of quantum statistics which includes the effects of mutual pairwise annihilation that is a defining feature of the \added{non-interacting} Majorana system.


\subsection{Outline of the present theory \added{and key differences between existing models}}

In this paper, we will address the problem of building a many-body theory of non-interacting Majorana fermions as Majorana first envisioned them: spin-$1/2$ neutral fermions identical to their own antiparticle state and that, therefore, exhibit a mutual pairwise annihilation. \added{In this way, we deviate strongly from the anyonic description, where the concept of individual, independent Majorana fermions is inherently unphysical.}

 \added{
A many-body theory of Majorana fermions beyond the traditional anyonic paradigm has garnered a large amount of interest in recent years 
in the form of the Sachdev-Ye-Kitaev (SYK) model, which consists of $N$ Majorana fermions $\chi_j$ interacting with a random, all-to-all four-point interaction $J_{ijk\ell}$}\cite{Kitaev_SYK,Maldacena,Polchinski}:
\begin{align}
\added{H_{SYK}=\sum_{i<j<k<\ell} J_{ijk\ell} \chi_i \chi_j \chi_k \chi_\ell}
\end{align}
\added{The parameters $J$ are pulled from a Gaussian distribution proportional to $N^3$, which ensures the interaction stays finite for large $N$:
}
\begin{align}
\added{J^2=\frac{N^3}{3!}\overline{J_{ijk\ell}^2}}
\end{align}
\added{Such long-range, random interactions allow us to describe the above as effectively a $0+1$ dimensional model of $N$ fermions on $N$ lattice sites, as there is no longer a concept of spatial distance. In this way, a large number $N>>1$ of lattice sites in SYK is synonymous with the large $N$ limit of Majorana degrees of freedom. Such a model was originally { introduced (for complex fields) in the study of nuclear many-body systems with simultaneous interactions, where eigenvalue densities have been shown to approach a Gaussian distribution for large particle number \cite{French_SYK,Mon,Brody,Pandey,Verb}. More recently, Hamiltonians exhibiting all-to-all random interactions have become of interest to condensed matter physicists, in particular in terms of a}
spin-S quantum Heisenberg magnet with Gaussian-random interactions (known as the complex SYK or cSYK\cite{Sachdev_Ye,Franz_BH}):}
\begin{align}
\added{H_{cSYK}=-\mu \sum_j c_j^\dagger c_j+\sum_{ijk\ell} J_{ijk\ell} c_i^\dagger c_j^\dagger c_k c_\ell}
\end{align}
\added{ In contrast to the above $H_{cSYK}$ model, we immediately see that the large $N$ limit of $H_{SYK}$ will continue to describe strong-coupling down to low energy due to the lack of a quadratic term.
%
%
%
In such a limit, the Majorana self energy can be solved exactly due to a dominant contribution from "melonic" diagrams in the perturbative expansion, which subsequently leads to a suppression of vertex corrections\cite{Maldacena,Fidel}. Also note that the SYK model has been shown to be maximally chaotic, in that its quantum Lyapunov exponent saturates to the maximal possible value $\Lambda_L=2\pi/\beta$\cite{Maldacena2}, where $\beta$ is the inverse temperature. 
 Following the work of Maldacena et. al., this leads us to conclude that the SYK model has the unique status of being an exactly-solvable toy model of an AdS black hole in the IR limit, where the system develops conformal symmetry\cite{Maldacena}. Moreover, as a direct consequence of the SYK model being maximally chaotic, the resistivity has been shown to scale linearly with temperature\cite{Xu}, mirroring the non-Fermi liquid behavior seen in the "strange metal" phase of the cuprate superconductors\cite{Casey}.
}

\added{
Despite the remarkable versatility of the SYK model and its applications to strongly correlated matter, its ability to describe a realistic analog of high-energy Majorana fermions in a condensed matter setting is severely limited. The direct synthesization of the SYK model via experimentally-available Majorana zero modes in Kitaev chains has faced serious obstacles, as the overlap between individual Majorana wavefunctions create an addition term in $H_{SYK}$ that is quadratic in the operators $\chi_j$. Although present in the cSYK model, bilinear terms destroy the strong-coupling behavior in the traditional formulation\cite{Franz_BH,Ehud}. Similarly, a completely random interaction described by values of $J$ drawn form a Gaussian distribution would have to be induced among the MZMs, otherwise the system is no longer exactly solvable. The most likely realization of the SYK model in a solid-state material might instead be realized by either coupling a large number of semiconducting quantum wires to a disordered quantum dot in 2D\cite{Chew} or by binding MZMs on the surface of a 3D topological insulator 
to a nanoscale hole threaded by magnetic flux quanta\cite{Pikulin}, but even these proposals are difficult to implement due to the unfeasibility of constructing a large array of semiconducting wires and a limited experimental understanding of the Fu-Kane superconductor. It appears then that the most promising avenue for building an SYK model in the lab might be to relax the conditions of all-to-all random interactions and the disappearance of a bilinear term, as suggested in \cite{Hurtubise}.}

\added{ Furthermore, even if we were to develop a perfect realization of the SYK model in the lab, any insight gained into a non-interacting gas of Majorana fermions is inconsequential. 
As stated before, the SYK model is not used to describe realistic Majorana physics; it is rather introduced as an exactly-solvable toy model for holographic black holes. This explains why the SYK model is usually described as a "black hole on a chip"\cite{Pikulin} as opposed to a "neutrino gas on a chip". { In addition, although random interactions are not necessarily important for the SYK model, such systems are usually plagued by some underlying condition that is not present in the non-interacting limit, such as coupling the fermions to massive scalar fields \cite{Takahiro} or replacing the the Gaussian interaction term with some real interaction proportional to a tensor model without quenched disorder\cite{Witten}.}
%
%
%
%
%
The "relaxed" variant of the SYK model given in \cite{Hurtubise} might describe a Majorana Hamiltonian with a purely bilinear term, yet even in this "non-interacting" model the Majorana wavefunctions must be considered to be randomly distributed in real and spin space. The resulting system is characterized by two phases: a gapped phase and a disordered Fermi liquid, neither of which resembling a completely non-interacting gas. Indeed, for any generalization of the SYK model, the mutual pairwise annihilation of Majorana fermions requires some four-fermion interaction, leading us to conclude that the SYK is unsuited to describe the effects of self-conjugation on a non-interacting Majorana gas.}

\added{
Yet another difficulty in extending the SYK to model "real" Majorana fermion systems is due to the fact that the SYK lives in $0+1$ dimensions. Higher-dimensional extensions have been made by coupling lattices of SYK clusters together with pair-hopping interactions\cite{Davison,Qi,Song}, although whether or not such $D>0$ generalizations support the desirable features of the $0+1$ dimensional model are highly questionable\cite{Khveshchenko}. Recently, a 2D analog of the SYK model has been proposed in the context of quantized Majorana fields\cite{Turiaci}, where the UV limit is described by $N$ copies of a topological Ising CFT and, once again, we face the same issues we had when considering the MZM system.}
%

\added{From the above discussion, there is a clear schism between what we call a "Majorana fermion" in the context of high-energy and condensed matter physics. }The main difference might be boiled down to our definition of the Majorana-like nature of a particle. The traditional condensed matter definition of a Majorana particle is a zero-energy mode described by second quantization operators that are the complex conjugate of one another. Such a system described by totally self-adjoint operators lacks a global $U(1)$ symmetry and, hence, also lacks a well-defined particle number or vacuum state. \added{The same is true for the operators $\chi_j$ in the SYK model and its higher-dimensional generalizations. Because a Hamiltonian defined by a completely self-adjoint set of fermionic operators automatically describes a topologically-nontrivial field that does not obey Pauli exclusion\cite{Wilczek_book}, a reliable analog of non-interacting, self-conjugate spin$-1/2$ fermions that obey CPT invariance is unlikely to be found in either the MZM or SYK systems.} 
 If we wish to consider a "true" condensed matter analog of Ettore Majorana's original idea, we must instead consider a non-interacting system defined by an initially anti-symmetric wave function that exhibits the possibility of symmetric correlation \added{through the relaxation of Pauli correlation} and, hence, \added{exhibits} mutual annihilation. \added{We expect that (in the limit of small to no coupling) the spin-statistics theorem will remain applicable, and therefore Majorana fermions will exhibit a stable ground state at $T=0$. This agrees with present calculations of high-energy Majorana fermions\cite{Dai} but directly contradicts the form of $H_{SYK}$, which is incompatible with bi-linear terms in the Hamiltonian proportional to a non-zero chemical potential and, subsequently, is always particle non-conserving for any temperature or interaction strength.} \added{It is for this reason that we can think of our model as a "number conserving" approximation for the Majorana system, where the total number of particles (fermionic$+$bosonic) is ultimately conserved.}
 
 \added{ In the context of condensed matter, a theory of Majorana fermions beyond a BCS mean-field description has been suggested as a necessary description for MZMs in p+ip superfluids, where the condensate might have interesting properties that could effect the traditional topological behavior undescribed in the usual Bogoliubov-deGennes mean-field theory\cite{Leggett_Lin1,Leggett_Lin2}. Raman spectroscopy studies on the Kitaev honeycomb RuCl$_3$ similarly appear to show conserved particle number in the many-body Majorana system\cite{Yiping}.}
%
%
%
%
%
\added{In the high-energy regime, this number-conserving approximation can be thought to be analogous to conservation of total momentum when two non-relativistic Majorana fermions\cite{Berg,Bern} (or even a positron/electron pair\cite{Klemperer, Arzimovitch}) annihilate. Such a model allows us to describe a non-interacting gas of Majorana fermions with standard thermodynamic arguments, leading to a closed form of the Majorana distribution function. This is impossible in the continuum limits of both the SYK and MZM systems, where a non-conservation of particle number implies long-range entanglement and hence a breakdown of freely-interacting, independent particle statistics\cite{Khare,Lerda}. In order to differentiate our model from that of the SYK or MZM models, we will refer to these "number-conserving" Majorana particles as Majorana-Schwinger fermions (MSFs), as we assume these particles are not anyonic and therefore obey the spin-statistics relationship, which was initially developed by Schwinger \cite{Schwinger} (see Table \ref{tab:table_1}). The resulting statistics will then be called Majorana-Schwinger (MS) statistics, to differentiate it from Fermi-Dirac, Bose-Einstein, or anyonic statistics.}

\begin{table}
\caption{\label{tab:table_1} \added{Comparison of the different properties of the Majorana zero mode (MZM), the Majorana fermions in the Sachdev-Ye-Kitaev model (SYK), the Majorana fermions in the higher dimensional extensions of the SYK model (Extended SYK), and the number-conserving Majorana-Schwinger fermions (MSF) considered in this work.}}
\begin{center}
\abovedisplayskip=0pt
\belowdisplayskip=0pt
\begin{tabular}{ |p{3cm}||p{3cm}|p{3cm}|p{3cm}|p{3cm}|  }
 \hline
  & MZM &SYK& Extended SYK & MSF\\
 \hline
 \hline
 Dimension   & $d=1,\,2$   &$d=0$&$d=1,\,2$ & $d=1,\,2,\,3$\\ \hline
 Interaction & Arbitrary  & All-to-all, \hspace{9mm} random & All-to-all, \hspace{9mm} random & Arbitrary \\
 \hline
 Conserved $U(1)$ & No & No & No & Yes \\ \hline 
 High-T limit & No & Yes & Yes & Yes \\ \hline
 Topologically trivial limit & No & Yes & Yes & Yes \\ \hline 
 Non-interacting limit & Yes, but still \phantom{....} entangled & No & No & Yes \\ \hline
\end{tabular}
\end{center}
\end{table}

In the \added{purely} statistical model \added{of MSFs} we will consider first, we assume mutual annihilation of particles whenever they occupy the same microstate. We only consider the fermionic degrees of freedom before going to the thermodynamic limit and observing the macroscopic, statistical effects of mutual annihilation. In the context of a microscopic Bose-Fermi \added{Hamiltonian}, we interpret the mutual annihilation we take for granted in our statistical model as a restriction on the Cooper instability in a 1D Bose-Fermi model exhibiting fermonic p-wave pairing, which we can than map to a free hopping model of \added{Majorana-Schwinger} fermions. By interpreting such annihilation as emergent bosonic behavior in the many-body system, we achieve a better approximation to the Majorana fermions discussed in the context of particle physics as opposed to the conventional MZM usually considered in topological media. We also obtain a conserved global $U(1)_\psi\times U(1)_\phi$ symmetry in such a model for fermions described by fermionic and bosonic fields $\psi$ and $\phi$, respectively, therefore permitting us to construct a well-defined vacuum state and particle number in the total Fermi-Bose system. As we will see in our microscopic model, the corresponding second-quantized Majorana\added{-Schwinger} operator will not be completely symmetric with respect to complex conjugation if we include an emergent bosonic component. Nevertheless, such an MSF formulation of the free Majorana gas will allow us to define physical observables with independent Majorana operators while still retaining the effects of \added{self-conjugation in the fermionic sector of the Hilbert space.}

We find that the many-body Majorana\added{-Schwinger} system exhibits bosonic statistics modulo-2, with the probability of two particles occupying the same quantum state now finite (as in the Bose-Einstein system) but with the number of possible states restricted to those with single or null occupation (as in the Fermi-Dirac system) due to particle-particle annihilation. \added{We continue to calculate the few-body configurational entropies of the system via combinatorial analysis. From a simple computational study, we propose a general form for the \added{MSF} entropy from which we derive \added{a closed form for the} Majorana\added{-Schwinger} distribution function, which illustrates a highly stable Fermi surface in the system.} Although a \added{similar low-temperature}, many-body theory of Majorana fermions has already been discussed as a bosonic extension of the Dirac negative energy sea \cite{Nielsen1, Nielsen2, Nielsen3}, such a study contradicts the accepted interpretation of a filled Dirac sea as the result of Pauli correlation, and is described via an unphysical interpretation of energy states \cite{Nielsen4}. Attempts to develop a Majorana equation of state are similarly plagued with unphysical analogies between the photon gas and the Majorana system \cite{Srivastava}. Our derivation of the Majorana\added{-Schwinger} statistics is based upon standard counting arguments used in the study of the fermionic system, and assumes nothing more than the basic assumptions of standard quantum statistical mechanics \cite{Fermi,Bedell1, Cowan}. 

The thermodynamics of the Majorana\added{-Schwinger} gas is \added{then} studied in depth in one, two, and three dimensions in the non-relativistic limit with a brief excursion into the 3D ultra-relativistic case, with clear differences and surprising similarities found between the \added{Majorana-Schwinger} and Fermi-Dirac systems. \added{The thermodynamics of the non-relativistic Majorana-Schwinger ensemble is then compared with that of other theoretical models and experimentally-studied materials that are proposed to harbor Majorana quasiparticles, with which we find a surprising amount of agreement in the limit where the effects of Pauli exclusion dominate mutual annihilation. Possible realizations of our MS statistics is then discussed in the context of astrophysical environments, where Majorana fermions might be present in high-density gases of extraterrestrial neutrinos.}

\section{II. \added{Boltzmann Entropy of the Majorana-Schwinger gas}}

\subsection{Signatures of a Fermionic ground state in a \added{non-interacting gas of Majorana fermions}}

In the development of a many body theory of the Majorana fermion, we face an immediate issue concerning the implications of the mutual pairwise annihilation that defines the Majorana system. It would appear that the closed system does not have a conserved number of particles, and that this might yield difficulties in the development of a statistical model. {When we develop the microscopic \added{realization} of our free Majorana\added{-Schwinger} system, we will deal with the non-conservation of particle number in more detail, but for now} we account for an apparent number-conservation violation by restricting our study to the grand canonical ensemble in the degenerate and thermodynamic limits. Similarly, such fluctuations in a conserved quantity as the number density can be thought to be analogous to the number fluctuations seen in Fermi liquids, which retain a constant total particle number in small subsystems of a larger system \cite{Swingle}. In the presence of these particle-number fluctuations, our system \added{simply} exhibits a larger quantity of microstate configurations compared to the traditional Fermi-Dirac system. Nevertheless, we face a greater issue if we consider the system to have statistical behavior dominated by mutual pairwise annihilation. If particle-particle annihilation dominates the Majorana statistics, then we will have strong variance of particle number about the mean even in the thermodynamic limit. This is in stark contrast to the fermionic system in the grand canonical ensemble, where variation in the mean particle number vanishes as we take the same limit. Moreover, of greater concern is the apparent impossibility of some non-bosonic Majorana ground state. It would then appear that, in the zero-temperature limit of the Majorana gas, all of the particles will favor annihilation and leave us with a ground state in the form of a photon gas. As such, there appears to be no viable \added{statistics describing the non-interacting, low-temperature limit} for the Majorana system, as the particles will immediately annihilate as soon as they begin to occupy the lowest energy level.

If we recall that the Majorana fermion is a spin$-1/2$ particle, then it should be clear that the many-body ground state is non-bosonic and is, indeed, identical to the case of a regular garden-variety fermion; i.e., the ground state of the Majorana\added{-Schwinger} gas should be a filled Fermi sea. By the spin-statistics theorem, the total wave function of the spin$-1/2$ Majorana fermion must be anti-symmetric. This is also seen in the anti-commutation relation $\{\gamma_i,\,\gamma_j\}=2\delta_{ij}$ that is satisfied by the Majorana operator $\gamma_j=\gamma_j^\dagger$. The Majorana fermion may experience a mutual pairwise annihilation, but such annihilation is impossible in a fully quantum mechanical description of the many-body theory. Turning back to the second quantized operator, it is often argued that $\gamma_j^2=1$ is the result of some finite probability of Pauli correlation "violation" in the Majorana system (similar to how $(c_k^\dagger)^2=0$ in the fermionic system implies a strict Pauli repulsion). We instead interpret this as a statement of {\it possible} annihilation of \added{particles in the Majorana-Schwinger formalism, where we assume the particles are completely non-interacting and therefore are not required to exhibit topological behavior of the MZMs}.  
Two fermions \added{at $T=0$} cannot occupy the same quantum state due to the anti-symmetric form of the many-particle wave function, and thus annihilation should not occur at zero temperature; i.e., whatever quantum statistics Majorana fermions follow should be equivalent to the Fermi-Dirac distribution in the ultra low-temperature regime.

To allow for mutual annihilation in the many-body Majorana system, we must consider the finite-temperature regime. As the thermal de Broglie wavelength decreases to smaller than the interparticle spacing, thermal effects dominate and the \added{fully-quantum mechanical} particles at zero temperature gradually lose their wave-like nature and begin to be described as \added{classical} particles obeying Boltzmann statistics. \added{In the language of the effective statistical potential, one might say (to a reasonable approximation) that the repulsive effects of Pauli exclusion tend to zero as room temperature is approached\cite{Mullin}.} As we increase temperature, we would therefore expect the Majorana\added{-Schwinger} statistics to be increasingly dominated by particle-particle annihilation, with the Majorana\added{-Schwinger} gas at high temperature to be \added{synonymous} a pure photon gas. Such suppression of the antisymmetric nature of the many-body spin$-1/2$ system is similarly considered in the path integral study of free Fermi gases, where the number of even permutations minus the number of odd permutations in a spin$-1/2$ ensemble (known as the fermion efficiency $\zeta$) approaches zero exponentially with decreasing temperature \cite{Ceperley1, Ceperley2, Ceperley3, DuBois}:
\begin{align}
\zeta=\exp\left(-2\beta N(\mu_F-\mu_B)\right)
\end{align}
where $\mu_F$ and $\mu_B$ are the fermionic and bosonic chemical potentials, respectively. It is therefore apparent that the negative contributions to some average observable $\langle \mathcal{O}\rangle$ resulting from antisymmetric particle permutations are minimized as we increase temperature. Therefore, in the case of free Majorana fermions \added{that respect the spin-statistics theorem}, we would expect that annihilation is gradually allowed as we increase temperature and "quench" the inherently antisymmetric character of the Majorana fermion system.

The dichotomy between anti-symmetric statistical correlation and mutual annihilation in the Majorana ground state is nothing new to our theory; many models of Majorana fermions in a cosmological setting consider the possible suppression of Pauli repulsion in the Majorana system in detail. A possible candidate for cold dark matter is a model consisting of the lightest neutralino, a popular candidate for the elusive WIMP (weakly interacting massive particle)\cite{Goldberg, Barger, Ellis}. Neutralinos are hypothetical Majorana fermions that form when the superpartners of the Z boson, the photon, and the neutral Higgs boson experience mixing from the effects of electroweak symmetry breaking \cite{Martin}. Due to Pauli correlation, the annihilation cross-section of neutalinos will become severely suppressed, resulting in a relic density of dark matter that exceeds current experimental observations and theoretical predictions from non-SUSY WIMPs \cite{Feng}. The exact relationship between Pauli repulsion and the annihilation cross-section in ultra-dense dark matter has been found explicitly by Dai and Stojkovic via a comparison of the mean free path for annihilation ($\lambda_a$) and the mean free path for the Pauli exclusion force ($\lambda_p$) \cite{Dai}. In regular dense Fermi matter, a degeneracy pressure builds as $\lambda_p$ shrinks to below the interparticle distance. In the neutralino system, however, Dai and Stojkovic find that this condition on the system is violated for high density; namely, the ratio $\lambda_p/\lambda_a\approx 1$ throughout the interior of a star made of pure dark matter. The authors conclude that neutralinos (and hence Majorana fermions in general) cannot follow regular Fermi-Dirac statistics due to dominating annihilation effects in the high density limit. This leads to a suppression of mutual annihilation in the low density limit and an abnormally high relic density of cold dark matter that exceeds present estimations based on the annihilation cross section of WIMPs \cite{Lee, Salati}. If we are to maintain agreement with experimental signals from high-energy gamma rays,
only low-density neutralino stars may exist \cite{Jie, Bi1, Bi2, Yury}. A reduced annihilation cross section also leads to better agreement with gravitational lensing observations of low density cores in triaxial halos of cold dark matter found in dwarf irregular galaxies \cite{Spergel, Tyson, Moore, Flores, Blok}.

Stojkovic and Dai's study 
implies that \added{the mutual annihilation of Majorana fermions that obey the fundamental laws of the Standard Model} is severely suppressed in the low-density, ultra low-temperature limit, and that, in this limit, the Majorana statistical distribution is identical to the regular Fermi-Dirac distribution. The goal of this paper is to derive the exact form of the non-interacting Majorana\added{-Schwinger} distribution function and see explicitly how the resulting statistics at zero and finite temperature differs from that of the traditional Fermi-Dirac system. Outside dark matter cosmology, the subtle interplay of particle-particle annihilation and Pauli exclusion in the Majorana system is often overlooked. For example, \added{a general} Majorana gas is often argued to have a chemical potential $\mu=0$ as a direct consequence of non-conservation of particle number, and unphysical similarities are often drawn between the bosonic photon gas and the fermionic Majorana system \cite{Srivastava}. Such an interpretation overlooks the anti-symmetric nature of the many-body Majorana system, and completely disregards the above studies on neutralino annihilation. Moreover, the non-conservation of particle number is not by any means a strong indicator of zero chemical potential. It can be shown that the chemical potential for light from non-incandescent sources may achieve a non-zero value, and in general the $\mu$ of a photon gas could take on any value due to reactions with collective excitations of matter  \cite{Wurfel, Herrmann, Landsberg}. The most straight-forward argument for $\mu=0$ in blackbody radiation is to build the distribution function from a microscopic argument and compare with the Bose-Einstein distribution \cite{Bai}. In a similar fashion, to build a statistical model of \added{our} Majorana\added{-Schwinger} fermions that correctly deals with particle annihilation, we must start from a microscopic counting argument and build the Majorana\added{-Schwinger} distribution without making any prior assumptions on the system. From the above arguments, there might be a fraction of the low-temperature Majorana\added{-Schwinger} gas that exhibits mutual pairwise annihilation as we raise the temperature, but there remains a fermionic component that does not exhibit this annihilation. The existence of this fermionic component will thus ensure a non-zero chemical potential in the entirety of the low-temperature system.

\begin{figure*}
\hspace{-25mm}\includegraphics[width=1.3\columnwidth]{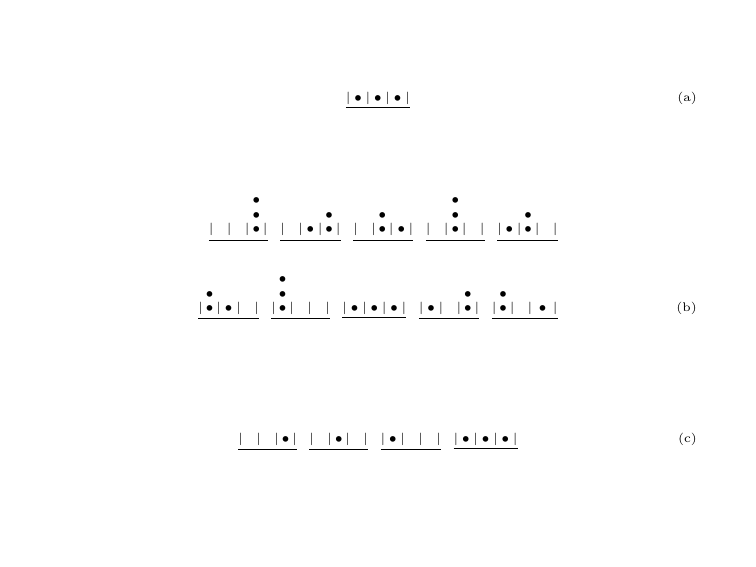}
\caption{\label{fig:pic2} An example of indistinguishable particle combinatorics for a simple $N=3$ spinless \added{Fermi-Dirac} (a), \added{Bose-Einstein} (b), and spinless \added{Majorana-Schwinger} (c) system with $G=3$ microstates. In the fermionic system (a), we are constrained to have only one possible configuration by the Pauli exclusion principle. In the bosonic system (b), we are not constrained by Pauli exclusion, and can therefore have a maximum of ten possible configurations. In the \added{Majorana-Schwinger} system (c), mutual particle-particle annihilation of identical particles with half-integer spin can be interpreted as a "violation" of the Pauli exclusion principle. This results in four possible configurations for the toy system above: the sum of the different possible configurations for one and three fermions.}
\end{figure*}

\subsection{Statistical weight of a Majorana\added{-Schwinger} gas from a modulo-2 variant of bosonic combinatorics}

In the vast majority of the present literature, researchers have tackled the many-body theory of Majorana-like particles by considering their exchange statistics. When confined to $2+1$ dimensions, the trajectories of quantum particles feature a one-to-one mapping with the elements of the braid group due to an arbitrary statistical phase under spatial exchange \cite{Leinaas, Wilczek_2}. The Hilbert space of a Kitaev chain or Kitaev honeycomb lattice in the Majorana representation then features a topological phase described by particles obeying Abelian or non-Abelian braiding statistics \cite{Kitaev, Kitaev_honey, Freedman}. In this work, however, we approach the many-body quantum statistics of a general $D+1$ dimensional Majorana\added{-Schwinger} fermion system by considering the effects of mutual pairwise annihilation on the traditional Fermi-Dirac combinatorics. \added{We assume "total" number conservation, dealing purely with the fermionic contribution to the Hilbert space for the time being. When we consider a microscopic Hamiltonian for our system later on in this work, we will explicitly show that the bosonic contribution is basically negligible in the low-temperature limit.} The microscopic description of our system is free from non-trivial topological effects, and is well-defined in the finite-temperature limit for any dimension (as opposed to the finite-temperature limit of two and three-dimensional systems harboring topological order) \cite{Preskill, Kitaev_finite_T, Chamon_2, Chamon_3, Ortiz}. Although the statistics of a simplified model of anyons has already been previously considered in the study of particles with "intermediate" exclusion statistics via a generalized Pauli exclusion principle  \cite{Haldane, Wu, Rachidi}, there has been a clear lack of discussion on the effects mutual annihilation has on the combinatorics of neutral, indistinguishable quantum particles in the topologically trivial regime. \added{A deviation from the braid group representation of low-dimensional MZMs (and instead towards a combinatorial interplay of annihilation and Pauli exclusion in a system of self-adjoint fermions) can then be considered to be the driving force behind this project.}\footnote{We thank Stefanos Kourtis for making this point.}

To fully understand the statistics of the \added{number-conserving Majorana-Schwinger ensemble}, we begin with a state-counting argument analogous to that of the fermionic system \cite{Cowan}. Recall from the Pauli exclusion principle that no two fermions with the same quantum numbers can occupy the same quantum state. The number of possible ways of arranging $N$ spinless fermions in $G$ microstates is subsequently given by $G$ choose $N$. This is in stark contrast to the bosonic system, where the number of possible configurations increases indefinitely with increased particle number. 
%
%

In the Majorana\added{-Schwinger} system, annihilation may be incorporated into the many-body statistics by considering all possible bosonic configurations for a system of size $N$ and disregarding all arrangements that harbor doubly-occupied states. The number of possible ways of arranging $N$ spinless \added{MSFs} in $G$ microstates is then the sum of distinct fermionic arrangements with an upper bound of $N$. This summation is only to be taken over configurations of an odd number of particles if $N$ is odd, and only over configurations of an even number of particles if $N$ is even. This is due to the annihilation of particles only affecting pairs of the same quantum state, leaving the remaining particles odd or even depending on the value of $N$. Hence, we can write the Majorana\added{-Schwinger} statistical weight as

\begin{align}
\Gamma=\begin{cases}
&\sum_{k\,\textrm{odd}}^N {G\choose k}, N\,\textrm{odd}\\\\
&\sum_{k\,\textrm{even}}^N {G\choose k},\,N\,\textrm{even}
\end{cases} \equiv \sum_{k}^N*{G\choose k}
\end{align} 

\noindent Unlike the fermionic case, the Majorana\added{-Schwinger gas} can support a many-body state with $N>G$. Due to pairwise annihilation, the statistical weight for this case will be equivalent to the weight for $N=G$ particles if $G-N$ is even and the weight for $N=G-1$ if $G-N$ is odd. This is a direct consequence of the modulo 2 bosonic behavior discussed earlier.

\begin{figure*}
\includegraphics[width=1\columnwidth]{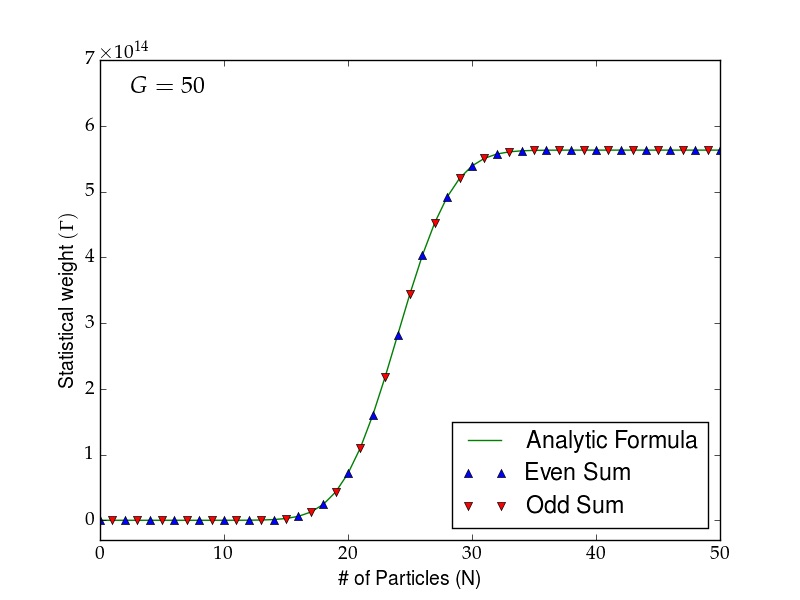}
\caption{\label{fig:pic1} The statistical weight for $N_j$ Majorana\added{-Schwinger} particles in $G_j=50$ microstates vs. $N_j$. The analytic formula of Eqn. \eqref{4} (solid green) is plotted alongside the partial binomial sum for even (blue triangle ) and odd (red triangle) values of different $N_j$. Such a plot gives us confidence in our derivation of the analytic formula.}
\end{figure*}

In Fig. \ref{fig:pic2}, we see the number of possible configurations for a system of three microstates and three \added{complex} fermions (a), three bosons (b), and three Majorana\added{-Schwinger} fermions (c). From the counting argument given above, we see that the allowed configurations in the Majorana\added{-Schwinger} system varies significantly from both the fermionic and bosonic systems. Nevertheless, on the surface of this argument, it appears that we are significantly overcounting the possible configurations for the Majorana\added{-Schwinger statistics}. This is due to an apparent confusion between pre-annihilation and post-annihilation number of the Majorana\added{-Schwinger} fermions; namely, \added{the incorrect view} that it is only the post-annihilation number of Majorana\added{-Schwinger} fermions that is a physical observable. 
Such an objection may be counteracted by considering how \added{the annihilation process} occurs in the many-particle Majorana\added{-Schwinger gas}. As discussed before, annihilation
is only possible in the finite temperature limit\added{, where the effective Pauli repulsion is reduced}. In a reasonably low-density \added{limit},
studies on neutralino systems have shown that Pauli repulsion will dominate the effects of mutual pairwise annihilation. To find out \added{the temperature-dependence of this annihilation}, we have to consider {\it all} possible configurations of the system, build the configurational entropy, minimize the thermodynamic potential, and build the temperature-dependent Majorana distribution function. Such analysis is identical to that used in the bosonic system if one wants to investigate the onset of Bose-Einstein condensation. To talk about pre- or post- annihilation in the Majorana\added{-Schwinger} system before we include the effects of temperature is analogous to considering pre- or post- condensation in the bosonic system before we build the Bose-Einstein distribution. As such, we do not overcount our possible configurations, and we may safely proceed to the derivation of the Majorana\added{-Schwinger} statistical weight before we can begin including the physical implications of mutual annihilation.

With the Majorana\added{-Schwinger} statistical weight defined \added{and justified}, it is now our goal to simplify the above value for $\Gamma$ in preparation for physical analysis of the configurational entropy. To do this, we consider the sum over $N_j$ particles and $G_j$ microstates in the $j$th group:
\begin{align}
\Gamma_j=\sum_k^{N_j}* {G_j \choose k}
\end{align}
For $G_j\approx N_j$, we utilize the expression for a general sum of binomial coefficients \cite{Koepf}. The restriction of the summation over even or odd values of $N_j$ can be taking into consideration by the addition or subtraction of an alternating binomial sum. Thus, if $G_j\approx N_j$, we can approximate the Majorana\added{-Schwinger} weight $\Gamma_j$ to go as a simple power of two:
\begin{align}
\sum_{k}^{N_j}*{G_j\choose k}&=2^{G_j-1}\approx 2^{N_j-1}\label{7}
\end{align} 
It is worth noting that, due to the above argument, the statistical weight of the $N_j=G_j$ Majorana\added{-Schwinger} system is equivalent to the weight of the fermionic system when $N_j=G_j/2$ in the thermodynamic limit. This is easily understood if we recall that the latter is effectively a system described by $G_j$ microstates with each microstate either being occupied or unoccupied. The Majorana\added{-Schwinger} ensemble in a "full" microstate configuration $N_j\ge G_j-1$ follows a similar description due to the possibility of particle-particle annihilation, except now the weight $2^{G_j}$ overcounts by a factor of two. We are therefore left with a statistical weight of $2^{G_j-1}$. In essence, as the number of \added{MSFs} in the system approaches the number of microstates, the statistics becomes identical to that of a two-level quantum system.

If we wish to consider the case of general particle number $N_j<G_j$, we may reformulate the partial sum of binomial coefficients in terms of a Gaussian hypergeometric function $_2 F_1 (1,\,N_j+1-G_j,\,N_j+2;\,-1)$. To incorporate the constraint of summation over even or odd values for $N_j<G_j$, we rewrite the alternating binomial sum in terms of a binomial coefficient times a factor of $(-1)^{N_j}$. Looking at the even contributions to this sum, we find that

\begin{align}
\sum_{k\,\textrm{even}}^{N_j}{G_j\choose k}&=\frac{1}{2}\left\{ 
\sum_{k}^{N_j}{G_j\choose k}+\sum_{k}^{N_j}(-1)^k{G_j\choose k}
\right\}\notag\\
&=2^{G_j-1}-\frac{1}{2}{G_j\choose N_j+1}
{}_2 F_1 (1,\,N_j+1-G_j,\,N_j+2;\,-1)+\frac{1}{2}{G_j-1\choose N_j}\label{2}
\end{align}

\begin{figure*}
\begin{subfigure}{.5\columnwidth}
\centering
\includegraphics[width=1.09\columnwidth]{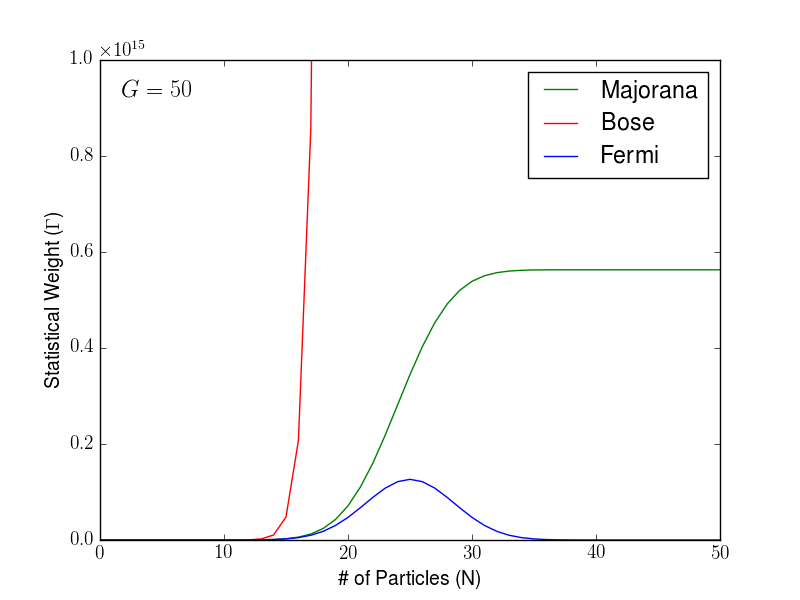}
\caption{}
\label{fig:sub1}
\end{subfigure}%
\begin{subfigure}{.5\columnwidth}
\centering
\includegraphics[width=1.09\columnwidth]{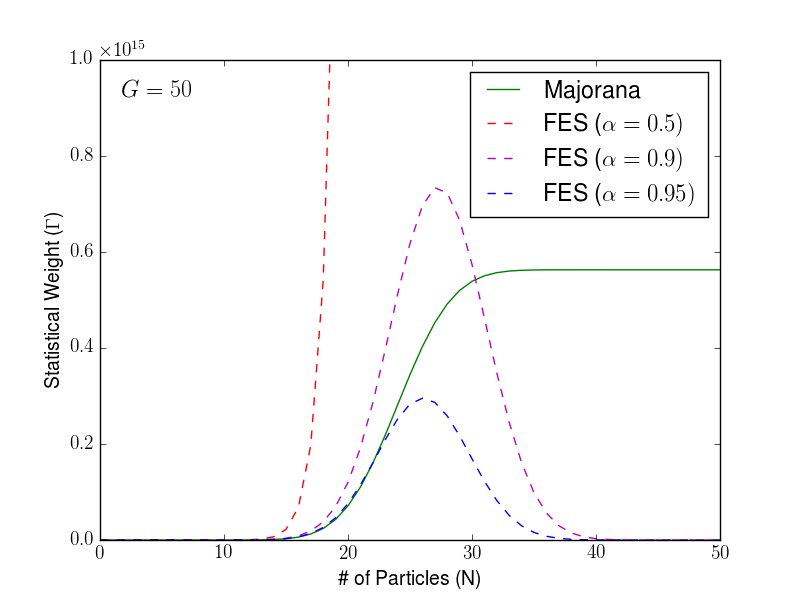}
\caption{}
\label{fig:sub2}
\end{subfigure}\\[1ex]
\caption{(a) The \added{Majorana-Schwinger} (green), \added{Bose-Einstein} (red) and \added{Fermi-Dirac} (blue) statistical weights vs. $N_j$ for  $G_j=50$. It appears that the \added{MSF} system is described by a completely different model of statistical mechanics from the regular bosonic and fermionic systems. However, from (b), we see that the Majorana\added{-Schwinger} system also differs from the "intermediate" statistics of Haldane and Wu for general $\alpha$. We are therefore left to conclude that the statistical mechanics of \added{number-conserving} Majorana fermions differ significantly from the statistical mechanics of particles with a conventional or generalized Pauli principle.}
\label{fig:test}
\end{figure*}

\noindent where, in the last line of the above, we have utilized the fact that $N_j$ is even to eliminate the $(-1)^{N_j}$ term. We proceed with an analogous calculation for the odd summation, which leads to a term identical to Eqn. \eqref{2}. This tells us that there is a single form for the Majorana\added{-Schwinger} statistics that is independent of whether or not the number of particles $N_j$ is odd or even. Simplifying the final term in Eqns. \eqref{2} by rewriting the binomial coefficient, the Majorana\added{-Schwinger} statistical weight can be written in the more concise form

\begin{align}
\Gamma_j&=2^{G_j-1}-\frac{1}{2}{G_j\choose N_j+1}\left\{{}_2F_1(1,\,N_j+1-G_j,\,N_j+2;\,-1)-\frac{N_j+1}{G_j}\right\}\label{4}
\end{align}

\noindent A plot of Eqn \eqref{4} vs. particle number $N_j$ is shown in Fig. \ref{fig:pic1} alongside the \added{MSF} $\Gamma_j$ in its discrete, summation form for both even and odd $N_j$.

\subsection{Comparison of the Majorana\added{-Schwinger} statistics with "intermediate" quantum statistics}

From Fig. \ref{fig:pic1}, it is clear that the statistical weight of a non-interacting gas of Majorana\added{-Schwinger} fermions deviates significantly from the regular fermionic weight. This is shown explicitly in Fig. \ref{fig:sub1}, where we have plotted the fermionic and bosonic weights alongside the \added{MSF result}. Such a plot illustrates the huge discrepancies between the \added{MS} many-body state and that of the Fermi and Bose systems, and hints that the former is an example of a completely new, distinct theory of quantum statistics.

Beyond the usual fermion or boson ensemble, it is also worth noting that the Majorana\added{-Schwinger} statistics varies significantly from the "intermediate" statistics that attempts to describe the many-body behavior of particles that interpolate between a fermionic and bosonic character. Often known as fractional exclusion statistics (FES), the theoretical groundwork for such a theory was first proposed by Haldane in 1991 and expanded upon by Y.S. Wu in 1994 \cite{Haldane, Wu}. The statistical weight of a gas described by the Haldane-Wu statistics is given by
\begin{align}
\Gamma_j={G_j+(N_j-1)(1-\alpha)\choose N_j}\label{11}
\end{align}
where the parameter $\alpha$ is defined as \cite{Khare}
\begin{align}
\alpha=-\left(
\frac{d_{N_j+\Delta N_j}-d_{N_j}}{\Delta N_j}
\right)
\end{align}
Here, $d_N$ is the dimension of the one-particle Hilbert space with the coordinates of all other $N_j-1$ particles held fixed and $\Delta N_j$ is the number of allowed changes to the particle number with fixed size and boundary conditions. Whereas $\alpha=0$ gives us bosonic statistics and $\alpha=1$ leads to fermionic behavior, the statistics of particles with arbitrary $\alpha\in (0,\,1)$ is known as parastatistics \cite{Green}. Unlike anyons, which are derived from the braid group and hence confined to two dimensions, parafermions and parabosons are based on the permutation group and can live in any dimension \cite{Lerda}. Although eqn. \eqref{11} faces difficulties in describing the free anyon gas (due to the fact that localized anyonic states lack nonorthogonality \cite{Leinaas, Wilczek_late}), we may still model the many-body anyon system with the above description if we assume a high magnetic field and very low temperature, thus confining the particles to the lowest Landau level \cite{Haldane, Murthy}. To a good approximation, the Haldane-Wu statistics described above \added{can be considered a new way} of looking at Abelian anyons through the lens of a generalized Pauli exclusion principle.

Statistical weights for the Haldane-Wu fractional statistics with $\alpha=0.5$, $0.9$, and $0.95$ are plotted in Fig. \ref{fig:sub2} alongside the Majorana\added{-Schwinger} weight. Much as in Fig. \ref{fig:sub1}, the intermediate statistics depicted in Fig. \ref{fig:sub2} bare little to no resemblance to that of the Majorana\added{-Schwinger} system. The differences between the Majorana\added{-Schwinger} statistics and the Haldane-Wu statistics is easily understood if we consider the microscopic foundations of the two theories. From the spin-statistics theorem, a model of quantum statistics that is "intermediate" between that of the Bose and Fermi systems must by described by particles which carry a spin "intermediate" between integer and half-integer values \cite{Rachidi}. As such, a system obeying FES is constructed by particles constrained by a generalized Pauli exclusion principle. The number of particles that are allowed to occupy the same quantum state (known as the "rank" of the parastatistics) vary depending upon the value of $\alpha$ \cite{Kaplan}. In constrast, Majorana\added{-Schwinger fermions} are defined as spin-$1/2$ fermions that 
\added{only begin to annihilate in the finite-temperature limit, when the effective statistical repulsion is suppressed}. 
The above gives us a clear conceptual difference between the anyonic/parafermionic and the MS ensembles, and supports the previous statement that the Majorana\added{-Schwinger} gas is described by a new theory of quantum statistics.

\subsection{Characteristics of the Boltzmann entropy for a Majorana\added{-Schwinger} gas of $N\approx G$ particles}

With a combinatorial formula for the Majorana\added{-Schwinger} $\Gamma_j$ now derived, we turn to evaluating the Boltzmann entropy for the system, given by

\begin{align}
S(N,\,G)=\sum_j \log (\Gamma_j(N_j,\,G_j))\label{1}
\end{align}
Due to the highly non-trivial form of the Majorana\added{-Schwinger} statistics, it is our present goal to simplify Eqn. \eqref{4} in a more digestible form that will allow us to better understand the underlying physics. For this purpose, we employ well-known identities for hypergeometric functions to transform ${}_2F_1(1,\,N_j+1-G_j,\,N_j+2;\,-1)$ in terms of a contour integral \cite{NIST}.

 We begin with the simplified case of $N_j\approx G_j$, and take $N_j=G_j-x$ where $x$ is some integer. Because the entropy of $N_j=G_j$ is trivial, it is a reasonable idea to begin with the case of $N_j\approx G_j$ to see the general behavior for smaller particle number.

\begin{table*}
\caption{\label{tab:table_eqn} Configurational entropy for a selected number of few-body Majorana\added{-Schwinger states}. From these examples, we can postulate an initial form for the configurational entropy at general particle number (see text).}
\centering
\abovedisplayskip=0pt
\belowdisplayskip=0pt
\begin{tabular}{ m{12.5cm} m{11cm} }
\hline\hline{\begin{subequations}\begin{flalign}
&S(N_j\ge G_j-1,\,G_j)=\sum_j \log (2^{G_j-1})\label{8a}&\\
&S(N_j=G_j-2,\,G_j)=\sum_j \log (2^{G_j-1}-1)\label{8b}&\\
 &S(N_j=G_j-3,\,G_j)=\sum_j \log (2^{G_j-1}-G_j)\label{8c}&\\
&S(N_j=G_j-4,\,G_j)=\sum_j \log \left(2^{G_j-1}-\frac{1}{2}(G_j^2-G_j+2)\right)\label{8d}&\\
&S(N_j=G_j-5,\,G_j)
=\sum_j \log \left(
2^{G_j-1}-\frac{1}{6}(G_j^{3}-3G_j^{2}+8G_j)\right)\phantom{H}
\label{8e}\\
&S(N_j=G_j-6,\,G_j)=\sum_j \log \left(
2^{G_j-1}-\frac{1}{24}(G_j^4-6G_j^3+23G_j^2-18G_j+24)
\right)\label{8f}&
\end{flalign}\label{NS_eq1}\end{subequations}}\\[0ex]
\hline\hline
\end{tabular}
\end{table*}

Starting with $N_j=G_j-1$, we find that the hypergeometric function in Eqn. \eqref{4} converges to unity. Eqn. \eqref{4} then tells us that, for $N_j=G_j-1$, the Majorana\added{-Schwinger} weight $\Gamma_j$ is given by a simple power of two:
\begin{align}
\Gamma_j(N_j=G_j-1,\,G_j)= 2^{G_j-1}
\end{align}
It is important to note that we have already seen, from Eqn. \eqref{7}, that $\Gamma_j$ follows an identical power law for $N_j=G_j$. From the discussion in the former section concerning the the case of $N_j>G_j$, it is now clear that the Majorana\added{-Schwinger} entropy Eqn. \eqref{1} remains linear with $G_j$ for all $N_j>G_j-1$.

 Proceeding to $N_j=G_j-2$, we follow the same procedure as for the $N_j=G_j-1$ case, and find that the hypergeometric function yields
\begin{align}
{}_2 F_1 (1,\,-1,\,G_j,\,-1)=1+\frac{1}{G}
\end{align}

\noindent The weight $\Gamma_j$ for $N_j=G_j-2$ is then given by $2^{G_j-1}-1$, from which the configurational entropy follows trivially. In this case, the entropy is nearly identical to the $N_j=G_j-1$ system, except now with a constant term subtracted from the power law.

Identical calculations give us the Boltzmann entropy for $N_j=G_j-3$, $N_j=G_j-4$, $N_j=G_j-5$, and $N_j=G_j-6$ Majorana\added{-Schwinger} particles. The results for all systems considered in this section are shown in Table \ref{tab:table_eqn}. From these expressions, it is reasonable to suggest that the entropy of a system of general particle number $N_j$ is given by

\begin{figure*}
\begin{subfigure}{.5\columnwidth}
\centering
\includegraphics[width=1.09\columnwidth]{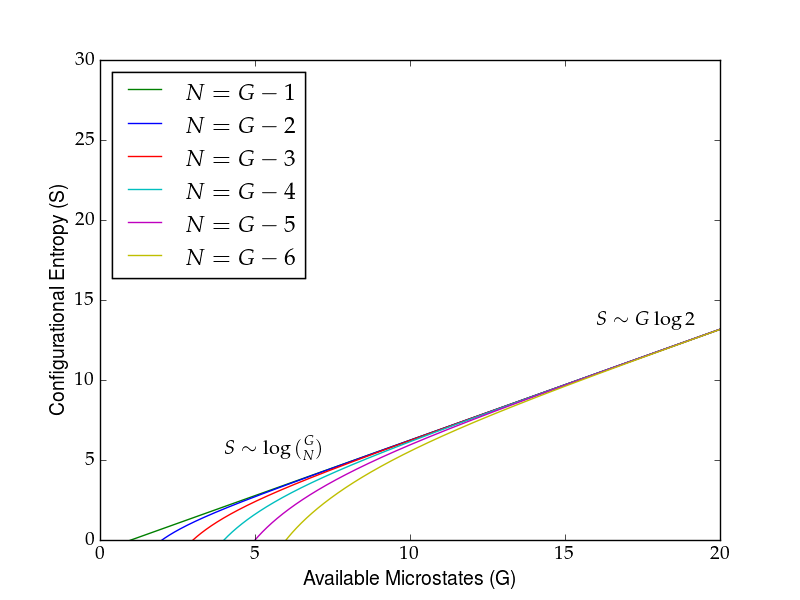}
\caption{}
\label{fig:sub3a}
\end{subfigure}%
\begin{subfigure}{.5\columnwidth}
\centering
\includegraphics[width=1.09\columnwidth]{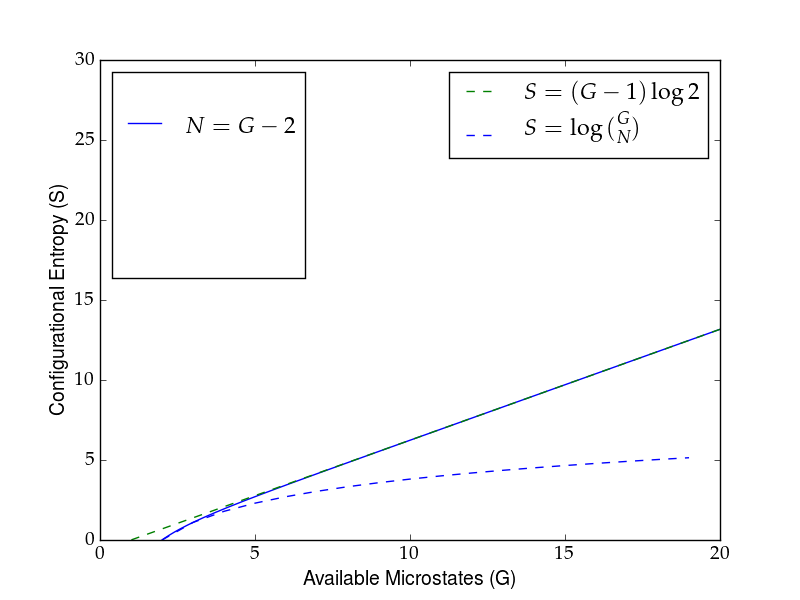}
\caption{}
\label{fig:sub3b}
\end{subfigure}\\[1ex]
\begin{subfigure}{.5\columnwidth}
\centering
\includegraphics[width=1.09\columnwidth]{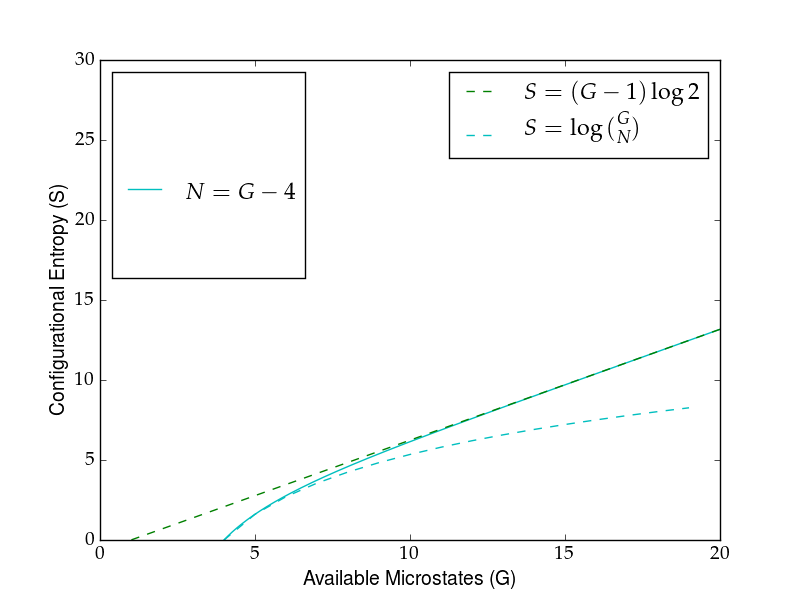}
\caption{}
\label{fig:sub3c}
\end{subfigure}%
\begin{subfigure}{.5\columnwidth}
\centering
\includegraphics[width=1.09\columnwidth]{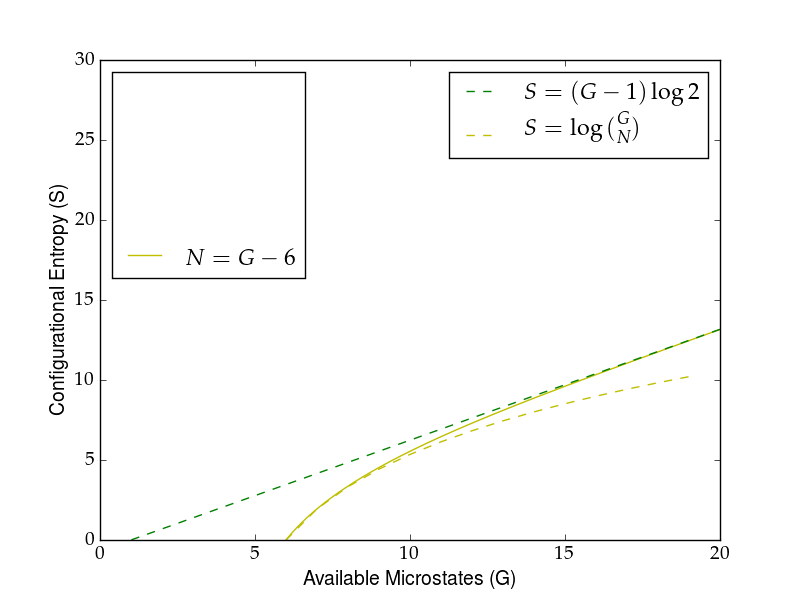}
\caption{}
\label{fig:sub3d}
\end{subfigure}\\[1ex]
\caption{(a) The configurational entropy of the Majorana\added{-Schwinger} system vs. the number of available microstates for $N\approx G$. For small $N$, we see the entropy starts out with fermionic behavior before converging to a universal value of $G\log 2$ in the large microstate limit. In (b)--(d), we explicitly see the fermionic behavior and subsequent transition to the "two-level" state for $N=G-2$, $N=G-4$, and $N=G-6$.}
\label{fig:4}
\end{figure*}

\begin{align}
&S(N,\,G)=\sum_j \log \bigg(
2^{G_j-1}-\frac{1}{(G_j-N_j-2)!}\sum_{k=0}^{G_j-N_j-2}\alpha_{k}^{(G_j-N_j-2)}G_j^k
\bigg)\label{9}
\end{align}

\noindent where $\alpha_k^{(G_j-N_j-2)}$ is some numerical constant dependent on $k$ and the upper bound $G_j-N_j-2$. Note that we define this coefficient such that it is zero for all values of $G_j-N_j-2<0$. It is interesting to note that the second term in the above bares a striking resemblance to the form of ${G_j\choose G_j-N_j-2}$ if we expand the binomial coefficient in terms of Stirling numbers of the first kind \cite{Graham, Knuth}.

With the Majorana\added{-Schwinger} weight and entropy now cast in a simpler form, we can easily analyze the system with $N_j<G_j-1$ particles. As we decrease the number of particles from the full or almost full state, the weight begins to decrease polynomially from that of the power of two behavior. The mediating term that reduces the number of possible states from the maximal "two-level" system is surprisingly fermion-like. It is worth wondering if, in some limit, the Majorana\added{-Schwinger} system exhibits the statistics of the regular fermion system. If we refer back to Fig. \ref{fig:sub1}, we indeed see that the Majorana\added{-Schwinger} weight approaches that of the fermionic system for low particle number. We can similarly turn to the entropies derived above to try and decipher if the Majorana\added{-Schwinger} system has fermionic-like behavior. In Figs. \ref{fig:sub3a}-\ref{fig:sub3d}, we plot the analytic formulae for the Majorana\added{-Schwinger} entropy given in Eqns. \eqref{8a}--\eqref{8f}. From these plots, it is clear that the Majorana\added{-Schwinger} entropy begins fermionic for small particle number and then approaches $(G_j-1)\log 2$ for larger values of $N_j$. We now turn to deriving a closed form for the Majorana\added{-Schwinger} entropy to analyze this fermionic behavior in greater detail.
\vspace{4mm}
\subsection{Closed form for the Majorana\added{-Schwinger} entropy at general particle number}

In order to derive the explicit form of the Majorana\added{-Schwinger} entropy for general particle number, recall the form of the statistical weight Eqn. \ref{4}. Now, we consider the case of $y=G_j-N_j$, where $y$ is an integer. Expressing the hypergeometric function in terms of a contour integral as we have done before, the Majorana\added{-Schwinger} statistical weight Eqn. \eqref{4} simplifies to

\begin{figure*}
\begin{subfigure}{.5\columnwidth}
\centering
\includegraphics[width=1.09\columnwidth]{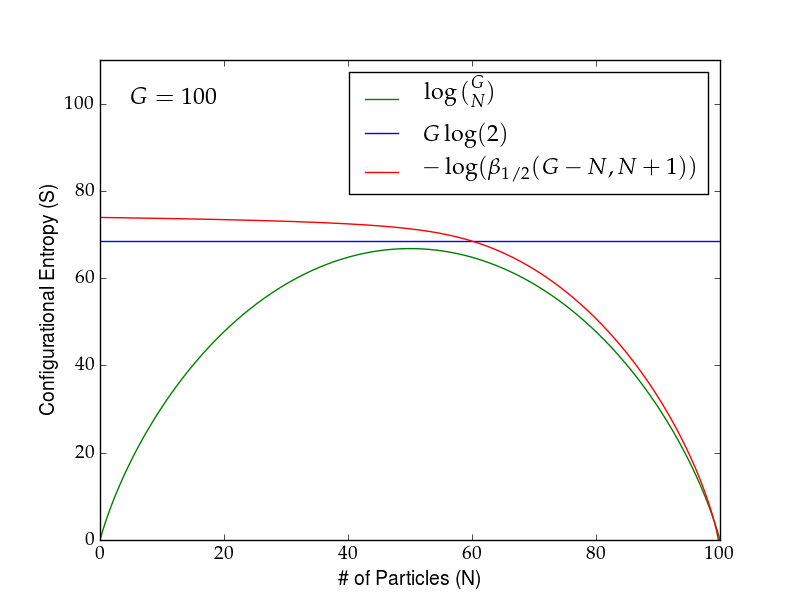}
\caption{}
\label{fig:sub4a}
\end{subfigure}%
\begin{subfigure}{.5\columnwidth}
\centering
\includegraphics[width=1.09\columnwidth]{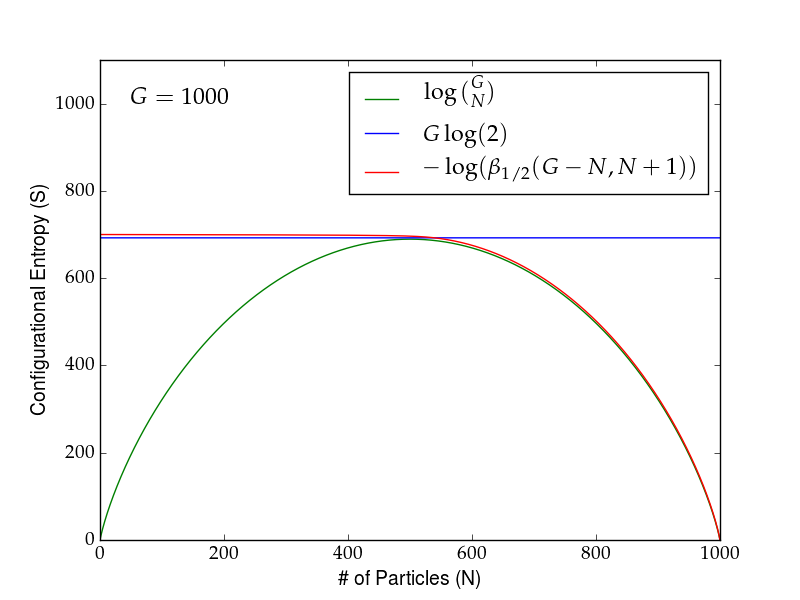}
\caption{}
\label{fig:sub4b}
\end{subfigure}\\[1ex]
\begin{subfigure}{.5\columnwidth}
\centering
\includegraphics[width=1.09\columnwidth]{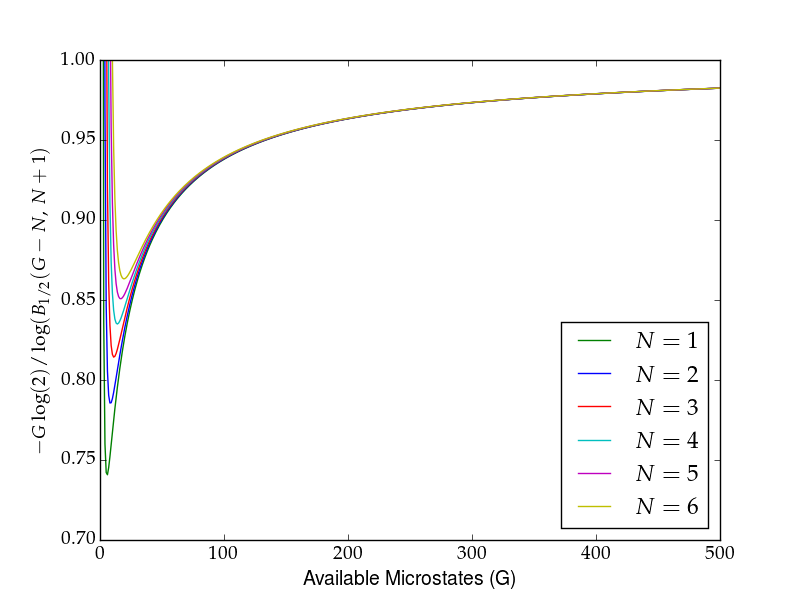}
\caption{}
\label{fig:sub4c}
\end{subfigure}%
\begin{subfigure}{.5\columnwidth}
\centering
\includegraphics[width=1.09\columnwidth]{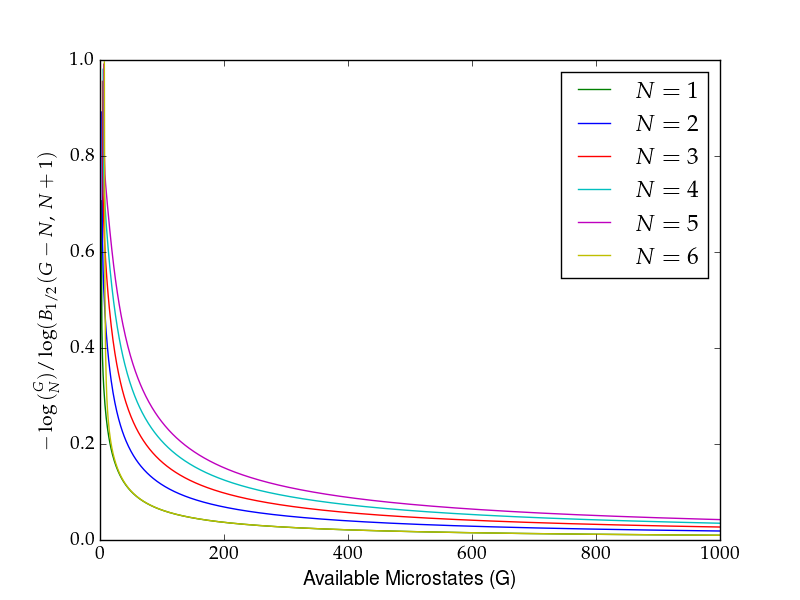}
\caption{}
\label{fig:sub4d}
\end{subfigure}%
\caption{(a) The components of the \added{Majorana-Schwinger} configuration entropy vs. number of particles for $G=100$. Shown are the fermionic (green), two-level (blue), and the negative of the beta function-dependent (red) components to the entropy. If we increase the number of microstates, as we can see in (b), the beta function term cancels out the $G\log 2$ at smaller particle number. As the number of particles increases, the beta function term cancels out the fermionic component. This effect can be seen in (c), where we have plotted the ratio of the linear-$G$ component and the beta function term for lines of constant particle number. As the number of microstates increases, the ratio approaches unity, as (a) and (b) appear to show. Similarly, (d) plots the ratio of the fermionic component and the beta function term for lines of constant particle number, showing a descend to zero for increased $G$.}
\label{fig:test}
\end{figure*}

\begin{align}
\Gamma_j&=2^{G_j-1}-\frac{1}{2}\left\{
\frac{1}{2}\textrm{Res}_1\left(
\frac{x^{G_j}}{(1-x/2)(1-x)^y}\right)-{G_j-1\choose G_j-y}
\right\}
\end{align}

\noindent The residue is significantly more complex then it was in the previous subsection. We lay out the derivation in Appendix A, the result of which is expressed in terms of the incomplete beta function $B_{1/2}(G_j-N_j,\,N_j+1)$:

\begin{align}
\Gamma_j
&\approx
2^{G_j-1}G_j{G_j-1\choose N_j}B_{1/2}(G_j-N_j,\,N_j+1)\label{13}
\end{align}

\noindent The Majorana\added{-Schwinger} configurational entropy for general particle number follows directly from the above:

\begin{align}
\ S(N,\,G)
\approx \sum_j G_j \log 2 +\sum_j \log {G_j\choose N_j}+\sum_j \log (B_{1/2}(G_j-N_j,\,N_j+1)\label{19}
\end{align}

From Eqn. \eqref{19}, we see that the configurational entropy of the Majorana\added{-Schwinger} system is composed of a term which is linear in $G_j$, a fermionic-type term, and a term dependent on the incomplete beta function. Instead of dealing with the beta function directly, we turn to simple numerics in order to understand what effects this function has on the physical behavior. In Figs. \ref{fig:sub4a} and \ref{fig:sub4b}, we plot the separate components of the configurational entropy for $G_j=100$ and $G_j=1000$, respectively. As we increase the microstates $G_j$, the negative of the log of the incomplete beta function cancels the linear-$G_j$ term for small particle number and cancels the fermionic term for larger particle number. This regulating behavior of the incomplete beta function term is seen more explicitly in Figs. \ref{fig:sub4c} and \ref{fig:sub4d}, where we plot the ratio of the linear $G_j$ component and the beta function term and the ratio of the fermionic component and the beta function term, respectively. With increased microstates, the former approaches unity and the latter approaches zero for fixed particle number, thus emphasizing the regulating nature of the beta function term.

From the above discussion, we can incorporate the behavior of the incomplete beta function via Heaviside theta functions:

\begin{align}
&S(N,\,G)\approx \sum_j \Theta(G_j/2-N_j)\log {G_j\choose N_j}\notag\\
&\phantom{S(N,\,G)\approx}+\sum_j \Theta(N_j-G_j/2)G_j \log 2\label{16}
\end{align}

\noindent The incomplete Beta function required for the description of the entropy in the presence of particle-particle annihilation can thus be eliminated in favor of a piecewise function for a different number of particles. For $N_j<G_j/2$, the \added{Majorana-Schwinger} entropy behaves fermionically, as can be seen in the examples of Fig. \ref{fig:4}. However, as we add more particles to the system while keeping the number of microstates constant, the entropy approaches a constant $G_j\log 2$ behavior, as can also be seen in Fig. \ref{fig:4}.\\
\vspace{.2mm}

\section{III. Derivation of the Majorana\added{-Schwinger} distribution function}

\subsection{Existence of a Fermi surface in a Majorana\added{-Schwinger} gas at finite temperature}

In the previous section, we examined in detail the combinatorics of the Majorana\added{-Schwinger} gas. Here, we examine the physical consequences of such a statistical theory. Our goal is to find a form of the Majorana\added{-Schwinger} distribution function for use in the development of the Majorana thermodynamics.

We begin by expressing Eqn. \eqref{16} in terms of the density $n_j=N_j/G_j$ and taking the continuum limit:
\begin{align}
S(N,\,G)
&\approx \sum_j \Theta(G_j/2-N_j)\left\{
G_j\log G_j-N_j\log N_j-(G_j-N_j)\log (G_j-N_j)
\right\}\notag\\
&\phantom{\approx}+\sum_j \Theta(N_j-G_j/2)G_j\log 2\notag\\
&=-\sum_j G_j\left\{
\Theta(1/2-n_j)\left\{
n_j \log n_j +(1-n_j)\log (1-n_j)
\right\}-\Theta(n_j-1/2)\log 2
\right\}\notag\\
&\rightarrow -V\sum_{p\sigma}\left\{
\Theta(1/2-n_{p\sigma})\left\{
n_{p\sigma}\log n_{p\sigma}+(1-n_{p\sigma})\log(1-n_{p\sigma})-\Theta(n_{p\sigma}-1/2)\log 2
\right\}
\right\}\label{26}
\end{align}

\noindent Minimizing the thermodynamic potential, we find the expression
\begin{align}
&\sum_{p\sigma}\left(
\epsilon_{p\sigma}^0-\mu +T \frac{ds}{dn_{p\sigma}}
\right)d n_{p\sigma}=0\label{17}
\end{align}
where $\epsilon_{p\sigma}^0$ is the interparticle energy, $\mu$ is the chemical potential, and $s$ is the thermodynamic entropy. Solving for $ds/dn_{p\sigma}$ yields 

\begin{align}
\frac{ds}{dn_{p\sigma}}&=-\sum_{p\sigma}\bigg(
-\delta(1/2-n_{p\sigma})\left\{
n_{p\sigma}\log n_{p\sigma}+(1-n_{p\sigma})\log (1-n_{p\sigma})\right\}\notag\\
&\phantom{=}+\Theta(1/2-n_{p\sigma})\log \left(\frac{n_{p\sigma}}{1-n_{p\sigma}}\right)-\delta(n_{p\sigma}-1/2)\log 2
\bigg)\notag\\
&=-\sum_{p\sigma}\Theta(1/2-n_{p\sigma})\log \left(\frac{n_{p\sigma}}{1-n_{p\sigma}}\right)
\end{align}

\begin{figure*}
\includegraphics[width=1\columnwidth]{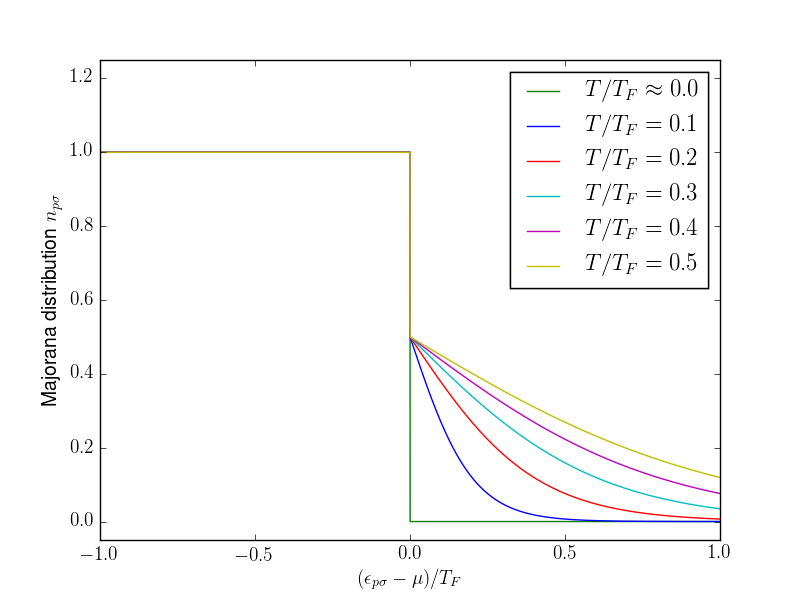}
\caption{\label{fig:pic7} The Majorana\added{-Schwinger} distribution function vs. energy for several values of temperature. No matter what temperature we consider, a "universal" discontinuity in the distribution remains.}
\end{figure*}

\noindent Plugging this into Eqn. \eqref{17}, we find the thermodynamic relation

\begin{align}
\epsilon_{p\sigma}^0-\mu+T \log \left(\frac{n_{p\sigma}^0}{1-n_{p\sigma}^0}\right)\Theta(1/2-n_{p\sigma})=0
\end{align}
Solving for the distribution function of the non-interacting Majorana\added{-Schwinger} gas $n_{p\sigma}$, we find the relation
\begin{align}
n_{p\sigma}^0=\frac{1}{\exp\left(\frac{\epsilon_{p\sigma}^0-\mu}{T\Theta(1/2-n_{p\sigma}^0)}\right)+1}\label{18}
\end{align}
Due to the Heaviside theta function, the above expression for the Majorana\added{-Schwinger} distribution function is self-consistent. However, we can significantly simplify the above if we consider the regions $n_{p\sigma}<1/2$ and $n_{p\sigma}>1/2$ separately. If we assume the former, then we obtain the normal fermionic distribution function. Because $n_{p\sigma}<1/2$ for $\epsilon_{p\sigma}^0-\mu>0$ in the fermionic system, it is easy to see that, above the Fermi surface, the Majorana\added{-Schwinger} distribution function behaves exactly like that of the fermionic. However, once $\epsilon_{p\sigma}^0-\mu<0$, the Majorana\added{-Schwinger} distribution function rises above a half, and the Heaviside theta function yields zero. This tells us that $n_{p\sigma}^0=1$ for all $\epsilon_{p\sigma}-\mu<0$, and we can thus rewrite Eqn. \eqref{18} in the more manageable form
\begin{align}
n_{p\sigma}^0&=\Theta(\mu-\epsilon_{p\sigma}^0)+\frac{1}{\exp\left(\frac{\epsilon_{p\sigma}^0-\mu}{T}\right)+1}\Theta(\epsilon_{p\sigma}^0-\mu)\label{27}
\end{align}
The distribution for several different temperatures is shown in Fig. \ref{fig:pic7}. This result is surprising, because it implies that there exists a sharp Fermi surface in the non-interacting Majorana\added{-Schwinger} gas even at finite temperature. Such a sharply defined Fermi surface is also seen in the non-interacting Fermi gas, but only at zero temperature. It follows from the discussion in the previous sections that this phenomenon is a direct consequence of the particle-particle annihilation within the Majorana\added{-Schwinger} system. The effects of such annihilation are encapsulated in the incomplete beta function term of the configurational entropy. It is also interesting to note that, from the form of Eqn. \eqref{27}, the statistics of the zero-temperature Majorana\added{-Schwinger} system is identical to that of the zero-temperature Fermi system, which agrees with previous studies on the Pauli exclusion of neutralino dark matter \cite{Dai}. Only as we increase temperature do we see a deviation from fermionic behavior in the Majorana\added{-Schwinger} system. 

\subsection{Dealing with the discontinuity at the Fermi surface}

Before we continue to the thermodynamics of the Majorana\added{-Schwinger} gas, it is important to first deal with the apparent discontinuity at the Fermi surface of the Majorana\added{-Schwinger} gas. For the purposes of this paper, we might ignore the sharp finite-temperature dip in the Fermi surface without any unwanted repercussions. However, such a discontinuity could prove to make the description of an interacting Majorana\added{-Schwinger} system, in which the Landau quasiparticles are only well-defined in the direct vicinity of the Fermi surface \cite{Pines}, somewhat problematic. We therefore briefly analyze the system near the Fermi surface here.

First, recall Eqn. \eqref{19} and take $N_j=G_j/2$. Using fundamental identities relating the incomplete and complete beta functions \cite{NIST}, the incomplete beta function in the above simplifies to a quotient of factorials in $G_j$ \cite{NIST}. The component of the entropy from the beta function term then yields
\begin{align}
\log \left(\frac{1}{2}\frac{(G_j/2)!^2}{G_j!}\right)&\approx-G_j \log 2
\end{align}
Thus, the total configurational entropy at $N_j=G_j/2$ appears to be completely fermionic. However, it is important to note that, as mentioned in an earlier section, the entropy of the fermionic system is identical to that of a two-level system:
\begin{align}
\log {G_j \choose G_j/2}&\approx G_j\log 2
\end{align}
We thus see that, in the close proximity to the Fermi level, we do not have a truly sharp discontinuity in the distribution function. Instead, we find a smooth transition between the $G_j \log 2$ and $\log {G_j\choose N_j}$ terms in the configurational entropy, which translates to a smooth transition of the distribution function at $\epsilon_{p\sigma}=\mu$. However, if we are not exclusively concerned with energy scales in the immediate neighborhood of the Fermi energy, we can assume the \added{Majorana-Schwinger} distribution function has a sharp discontinuity at finite temperature without issue.

{ \subsection{Majorana\added{-Schwinger} statistics from a microscopic Fermi-Bose Hamiltonian}

In a gas of \added{number-conserving} Majorana fermions such as those considered above, we find a dominant Pauli repulsion that "protects" the system against complete annihilation in the ground state. However, we cannot appeal to such an argument if we wish to consider a \added{general} gas of Majorana particles found in condensed matter systems. The spin-statistics theorem (on which our previous argument relies) is not obeyed on discrete lattice systems due to a violation of Lorentz invariance \cite{Duck, Iaizzi}. This is seen in the case of bosonic spinons in frustrated quantum antiferromagnets \cite{Wen1, Scammell}. For this reason, if we are to analyze the low-temperature behavior of the free Majorana\added{-Schwinger} gas in a condensed matter scenario, we should consider model Hamiltonians that exhibit Majorana quasiparticles as fundamental excitations, and see explicitly if fermionic behavior is present in their ground state. Unlike regular fermions, these Majorana zero modes (MZMs) are equal superpositions of two complex fermions of equal spin. We can therefore think of a Majorana zero mode as the two real "halves" of a complex fermion. This is apparent if we decompose the complex fermionic operators $c_j$ and $c_j^\dagger$ into two Majorana operators $\gamma_j^{(1)}$ and $\gamma_j^{(2)}$:
\begin{subequations}
\begin{equation}
c_j^\dagger =\frac{1}{2}(\gamma_j^{(1)}+i\gamma_j^{(2)})
\end{equation}
\begin{equation}
c_j=\frac{1}{2}(\gamma_j^{(1)}-i\gamma_j^{(2)})
\end{equation}
\end{subequations}
These operators obey the anticommutation relation $\{\gamma_i,\,\gamma_j\}=2\delta_{ij}$ as discussed earlier. To find a system in which we can create unpaired Majorana zero modes, let us consider the spinless 1D superconducting Kitaev chain exhibiting triplet (p-wave) pairing, given by the Hamiltonian \cite{Kitaev}

\begin{align}
H(t,\,\mu,\,\Delta)&=-\mu \sum_{j=1}^N c_j^\dagger c_j-t\sum_{j=1}^{N-1} \left(c_j^\dagger c_{j+1}+h.c.\right)
+\Delta \sum_{j=1}^{N-1} (c_j c_{j+1}+h.c.)
\end{align}
We can see that, for $\Delta=t=0$ with $\mu<0$ (the topologically trivial phase), the above reduces to the Hamiltonian for free complex fermions. In other words, the MZMs on adjacent lattice sites are coupled. However, for $\Delta=t\not=0$ and $\mu=0$ (the topologically non-trivial phase), we find that the MZMs on alternating lattice sites are coupled, leading to unpaired zero modes at the ends of the wire. Below are the Hamiltonians for the topologically trivial and non-trivial phases, respectively:

\begin{subequations}
\begin{equation}
H(t=0,\,\mu<0,\,\Delta=0)=-i\frac{\mu}{2}\sum_j^{N-1} \gamma_j^{(1)}\gamma_j^{(2)}
\end{equation}
\begin{equation}
H(t\not=0,\,\mu=0,\,\Delta=t)=-it \sum_{j=0}^{N-1}\gamma_j^{(1)}\gamma_{j+1}^{(2)}
\end{equation}
\end{subequations}

In this paper, we wish to study the statistical and thermodynamic behavior of a free Majorana\added{-Schwinger} fermion gas. However, the gas of Majorana zero modes considered in the cases above are, as mentioned before, not independent particles free of their parent fermion. This is apparent when one attempts to write an \added{MZM} number operator, which can only be written in terms of the fermionic number operator:
\begin{align}
c_j^\dagger c_j=1-i\gamma_j^{(1)}\gamma_j^{(2)}
\end{align}
In other words, we need two Majorana zero modes $\gamma_j^{(1)}$, $\gamma_j^{(2)}$ (i.e., a single complex Dirac fermion) to define a physical observable. To define a Hamiltonian which exhibits a localized Majorana mode as an elementary excitation, we have to therefore break a $U(1)$ symmetry and subsequently lose a well-defined particle number \cite{Hassler}. It is then clear that we are unable to define a free gas of independent MZMs described by $\gamma_j^{(1)}$ and $\gamma_j^{(2)}$ on the Kitaev chain. The Kitaev chain also faces difficulties in our study due to the fact that the unpaired MZMs that appear at the ends of the Kitaev chain obey highly non-trivial non-Abelian exchange statistics \cite{Read, Ivanov, Rao}. In this study, we wish to consider the simpler case of a gas of free fermions which are their own antiparticle, and it is for the reasons above that we should consider a different toy model than the 1D superconducting chain first proposed by Kitaev. 

To consider a possible microscopic realization of our Majorana\added{-Schwinger} statistics and the Majorana\added{-Schwinger} thermodynamics, let us recall our previous argument concerning the possibility of mutual annihilation in the \added{MSF} gas. When two Majorana fermions annihilate, they cannot be described by a completely antisymmetric many-body wave function, whether or not they are integer or half-integer spin. It is for this reason that \added{number-conserving} Majorana fermions in the form of neutrinos or neutralinos cannot annihilate at zero temperature. In our solid state system, we can include the effects of mutual annihilation by extending the Hilbert space \added{of the traditional MZM} to include both fermionic and bosonic contributions. This is done by introducing the Majorana\added{-Schwinger} fermion operators $\widetilde{\gamma}_j^\dagger $ and $\widetilde{\gamma}_j$, given by

\begin{align}
\widetilde{\gamma}_j^\dagger =c_j^\dagger +c_j b_j^\dagger b_j^\dagger ,\qquad \widetilde{\gamma}_j=c_j+c_j^\dagger b_j b_j
\end{align}

\noindent where $c_j^\dagger,\quad c_j$ are fermionic operators and $b_j^\dagger,\,b_j$ are bosonic operators. The effect on the fermionic Hilbert space is identical to that of Kitaev's MZM operators, however we now have a bosonic contribution that must be present if we are to consider the effects of emergent annihilation \added{while simultaneously replicating a "spin-statistics" relation (i.e., photon creation)}. Although this contribution is not needed in the context of Kitaev's original formulation of the Majorana zero mode, the inclusion of a bosonic extension 
in the Hilbert space is required to study the thermodynamic signatures of mutual annihilation in a microscopic formulation of our free Majorana\added{-Schwinger} gas. It is interesting to note that the anti-commutation relations of the modified Majorana\added{-Schwinger} operators $\widetilde{\gamma}_i$ and $\widetilde{\gamma}_i^\dagger$ have fermionic and bosonic-like features:

\begin{subequations}
\begin{equation}
\{\widetilde{\gamma}_i,\,\widetilde{\gamma}_j^\dagger \}=\delta_{ij}\left\{
1+2c_i^\dagger c_j(1+2b_j^\dagger b_i)+b_j^\dagger b_j^\dagger b_i b_i
\right\}
\end{equation} 

\begin{equation}
\{\widetilde{\gamma}_i,\,\widetilde{\gamma}_j\}=\delta_{ij}(b_j b_j+b_i b_i)
\end{equation}

\begin{equation}
\{\widetilde{\gamma}_i^\dagger,\,\widetilde{\gamma}_j^\dagger\}=\delta_{ij}(b_j^\dagger b_j^\dagger +b_i^\dagger b_i^\dagger)
\end{equation}

\end{subequations}
\noindent For $i\not=j$, the anti-commutator is zero, while for $i=j$ bosonic like behavior emerges unseen in the traditional Kitaev-Majorana formulation. Similar behavior is also seen in the commutation relations (see Appendix B for more details).

\added{The above} formulation allows us to define a coherent number operator for an arbitrary number of free Majorana\added{-Schwinger} fermions in terms of fermionic and bosonic operators:
\begin{align}
\widetilde{\gamma}_j^\dagger \widetilde{\gamma}_j&=c_j^\dagger c_j+c_jc_j^\dagger b_j^\dagger b_j^\dagger b_j b_j\notag\\ 
&\notag\\ 
&= c_j^\dagger c_j-b_j^\dagger b_j\left(
b_j^\dagger b_j-1
\right)\left(c_j^\dagger c_j-1\right)
%
\end{align}
If the $j$-th site is occupied by \added{an MSF} pre-annihilation, then $\widetilde{\gamma}_j^\dagger \widetilde{\gamma}_j=c_j^\dagger c_j$. 

Now that we have defined a modified Majorana operator that includes the effects of mutual annihilation, we need to consider a Hamiltonian that can be mapped to a free Majorana\added{-Schwinger} Hamiltonian. Let us start by considering the following general Fermi-Bose Hamiltonian:

\begin{align}
\widetilde{H}(t,\,\mu,\,U,\,V)&=-\mu \sum_{j=1}^N c_j^\dagger c_j -t\sum_{j=1}^N \left(c_j^\dagger c_{j+1}+h.c.\right)+V\sum_{j=1}^N \left(
c_j^\dagger c_{j+1} b_{j+1}^\dagger b_{j+1}^\dagger b_j b_j +h.c.
\right)\notag\\
&\phantom{=}+U\sum_{j=1}^N \bigg(
c_j^\dagger c_j c_{j+1}^\dagger c_{j+1}
\bigg\{
b_{j+1}b_{j+1}b_{j+1}^\dagger b_{j+1}^\dagger +b_j b_j b_j^\dagger b_j^\dagger\bigg\}
\bigg)
\end{align}
The first two terms in the above are identical to the first two terms in Kitaev's spinless 1D model. The third term describes the exchange of a single fermion with a pair of bosons from the $j$th site to the $j+1$st site, while the fourth term describes a nearest-neighbor fermion pair interaction that is amplified in the presence of bosons. We can interpret such a Hamiltonian to describe a tight-binding chain of 1D spinless fermions which interact with a Bose-Hubbard system in the $n=2$ Mott insulating phase or in a pair superfluid phase \cite{Valiente, Mishra, Stefan, Jiang}.

To represent the above in terms of the free Majorana\added{-Schwinger} representation, we first perform a mean-field approximation on the fourth term, taking $\Delta\equiv U\langle c_{j}^\dagger c_{j+1}^\dagger b_{j+1}b_{j+1} \rangle=U\langle c_{j+1} c_j b_{j+1}^\dagger b_{j+1}^\dagger \rangle=U\langle c_{j+1}^\dagger c_{j}^\dagger b_j b_j\rangle =U\langle c_{j}c_{j+1} b_{j}^\dagger b_{j}^\dagger \rangle$ as the order parameter of our theory:
\begin{align}
&Un_j^{(f)}n_{j+1}^{(f)} \left\{
b_{j+1}b_{j+1}b_{j+1}^\dagger b_{j+1}^\dagger +b_j b_j b_j^\dagger b_j^\dagger
\right\}\notag\\
=&U \left\{c_j^\dagger c_{j+1}^\dagger b_{j+1} b_{j+1}c_{j+1} c_j b_{j+1}^\dagger b_{j+1}^\dagger+c_{j+1}^\dagger c_{j}^\dagger b_j b_j c_{j}c_{j+1} b_j^\dagger b_j^\dagger \right\}\notag\\
\approx&\Delta \bigg\{ c_j^\dagger c_{j+1}^\dagger \left(b_{j+1} b_{j+1}-b_j b_j\right)+c_{j+1}c_j \left(b_{j+1}^\dagger b_{j+1}^\dagger-b_j^\dagger b_j^\dagger \right) \bigg\}
\end{align}

{ It is important to note here that such a mean-field expansion is equivalent to defining a restrictive condition on p-wave Cooper pairing in our system. Fermions on adjacent sites will have a non-zero pairing contribution only if there is an imbalance of bosons on these sites. Also note that our mean-field expansion preserves a global $U(1)_\psi\times U(1)_\phi$ symmetry, where the $U(1)_\psi$ symmetry is described by $\psi(x)\rightarrow e^{i\theta_\psi}\psi(x)$ for a fermionic field $\psi(x)$ and a $U(1)_\phi$ symmetry is described by $\phi(x)\rightarrow e^{i\theta_\phi}\phi(x)$ for a bosonic field $\phi(x)$. The $U(1)_\psi\times U(1)_\phi$ symmetry of the composite Fermi-Bose system described by $\psi^*(x) \phi(x)\rightarrow e^{i(\theta_\phi-\theta_\psi)}\psi^*(x)\phi(x)$ is then conserved if we take $\theta_\psi=\theta_\phi$. Such a symmetry is similarly seen in a Bose-Fermi realization of a two-channel model of Feshbach resonance \cite{Ludwig}. This is highly different from a purely fermionic Majorana \added{model}, where the particles described by the Eddington-Majorana equation break a global phase with the inclusion of a mass term, and are, hence, completely neutral \cite{Duff}. This is also different from Kitaev's original formulation, where a Hamiltonian with localized Majorana zero modes breaks a $U(1)$ symmetry from a mixing of the two modes $\gamma_j^{(1)}$ and $\gamma_j^{(2)}$. In our low-temperature system, however, mutual annihilation cannot occur for a system described by a purely anti-symmetric wave function, and thus we must consider an emergent symmetric contribution which subsequently permits the emergent composite particles to couple to a $U(1)$ gauge potential \footnote{We thank Debanjan Chowdhury for making this point.}. To \added{simultaneously} include mutual annihilation and a well-defined particle number in a condensed matter realization \added{of self-conjugate fermions}, we must forgo absolute neutrality of our composite particles. Nevertheless, even though the composite Bose-Fermi pairs are not neutral, the effect on the fermionic portion of the Hilbert space remains the same as that of the Kitaev-Majorana formalism; i.e., the fermionic component remains \added{self-conjugate. In this way, we might say that the Majorana-Schwinger representation weakly "sacrifices" absolute self-conjugacy of the Majorana fermions in favor of building a closed form of the statistical distribution function.} The bosonic restriction on p-wave Cooper pairing and the subsequent coupling of each neutral Majorana fermion to a bosonic pair allows us to study the mutual annihilation hinted at in cosmological and subatomic particle phenomenon in a coherent theory of quantum statistics. It is also worth noting that the absolute neutrality of a \added{fundamental, Standard Model} Majorana fermion is a sufficient but not necessary condition, as the Majorana fermion might interact via some electromagnetic toroidal dipole moment or some "hidden" $U(1)$ gauge interaction considered in the context of atomic dark matter \cite{Anapole, Fitzpatrick, Rosetti, Feldman, Cline}.

We can interpret the above condition for a conserved $U(1)_\psi\times U(1)_\phi$ symmetry by writing down the time derivative of these phases in terms of the fermionic and bosonic components' chemical potentials\footnote{Such a derivation is analogous to Feynman's explanation of Josephson tunneling in a superconducting junction \cite{Feynman}.}:

\begin{align}
-\hbar \frac{\partial}{\partial t}(\theta_\psi-\theta_\phi)=\mu_\psi-\mu_\phi
\end{align}
\\
We can easily see that the condition for a conserved global symmetry is that the bosonic and fermionic components are in thermal equilibrium with each other--i.e., $\theta_\psi=\theta_\phi\rightarrow \mu_\psi=\mu_\phi$. We have seen that in the statistical model (and, as we will soon see, in the microscopic model as well) that the emergent bosonic degrees of freedom are strictly temperature-dependent, and as such a thermal equilibrium between the fermionic and bosonic fractions in our system can be realized if we simply hold temperature constant. If such a condition is met, then we preserve a global $U(1)$ symmetry, and thus we retain a well-defined particle number and a well-defined Fermi surface. This is in sharp contrast to the traditional s-wave or p-wave Cooper pairing mechanism, in which the $U(1)$ symmetry of the free particle system is spontaneously broken down to $\mathbb{Z}_2$ in the mean-field expansion; i.e., $\theta_\psi\in \{0,\,\pi\}$, \added{which leads} to a Hamiltonian which does not conserve particle number. 
 }

Taking \added{the above} mean field expansion, the fourth term in the \added{modified Kitaev} Hamiltonian $\widetilde{H}$ defined above corresponds to a modified p-wave superconducting term. In such a system, the creation of a p-wave pair of fermions is directly dependent on the relative concentration of bosons on the adjacent sites; i.e., adjacent fermions will pair only if there is an imbalance of bosons on these sites. Taking $V=t$, we can write down the Majorana\added{-Schwinger} Hamiltonian in the mean field approximation:
\begin{align}
\widetilde{H}_{MF}(t,\,\mu,\,\Delta)&=-\mu \sum_{j=1}^N c_j^\dagger c_j -t\sum_{j=1}^N \left(
c_j^\dagger c_{j+1}\left\{1-b_{j+1}^\dagger b_{j+1}^\dagger b_j b_j\right\}+h.c.
\right)\notag\\
&\phantom{=}+\Delta \sum_{j=1}^N \left(
 c_jc_{j+1} \left(b_j^\dagger b_j^\dagger-b_{j+1}^\dagger b_{j+1}^\dagger\right)+h.c.
\right)
\end{align}
Note that this is nearly identical to the original Kitaev spin chain, except now we have additional terms corresponding to fermionic/bosonic exchange (in the hopping term) and a bosonic component in the superconducting term. Taking $\Delta =-t$, we are able to map our mean-field Hamiltonian to that of a free Majorana\added{-Schwinger} model defined by the operators previously introduced:
\begin{align}
\widetilde{H}_{MF}(t,\,\mu=0,\,\Delta=-t)&=t\sum_j \left(\widetilde{\gamma}_j^\dagger \widetilde{\gamma}_{j+1}+h.c.\right)
\end{align}
Here, we take $\mu=0$ for simplicity. 

Now that we have a microscopic Hamiltonian which can be mapped to the free Majorana\added{-Schwinger} system described by $\widetilde{\gamma}_j^\dagger$, $\widetilde{\gamma}_j$, we can now see if such \added{a construction yields}
thermodynamic observables similar to our statistical model. To do this, we first move our Majorana\added{-Schwinger} Hamiltonian to Fourier space:
\begin{align}
\widetilde{H}_{MF}&=\sum_k \xi(k) \widetilde{\gamma}_k^\dagger \widetilde{\gamma}_k
\end{align}
where $\xi(k)=2t\cos(k)-\mu$ for some general chemical potential. The term in the sum above is just the Majorana\added{-Schwinger} number operator in k-space. The expectation value of the number operator is readily found from the form derived beforehand, and can be expressed in terms of the fermionic and bosonic distribution functions $n^{(f)}(\xi_k)$ and $n^{(b)}(\xi_k)$, respectively:
\begin{align}
 \langle \widetilde{\gamma}_k^\dagger \widetilde{\gamma}_k \rangle
&=\frac{\textrm{Tr}\left(e^{-\beta \widetilde{H}_{MF}} \widetilde{\gamma}_k^\dagger \widetilde{\gamma}_k\right)}{\textrm{Tr}(e^{-\beta \widetilde{H}_{MF}})}\notag\\
&=\frac{\textrm{Tr}\left(e^{-\beta \widetilde{H}_{MF}}(c_k^\dagger c_k+b_k^\dagger b_k (b_k^\dagger b_k-1)(1-c_k^\dagger c_k)\right)}{\textrm{Tr}(e^{-\beta \widetilde{H}_{MF}})}\notag\\
&=n^{(f)}(\xi_k)+n^{(b)}(\xi_k)\left(
n^{(b)}(\xi_k)-1
\right)\left(
1-n^{(f)}(\xi_k)
\right)\label{micro}
\end{align}

\begin{figure*}
\begin{subfigure}{.47\columnwidth}
\centering
\includegraphics[width=1.09\columnwidth]{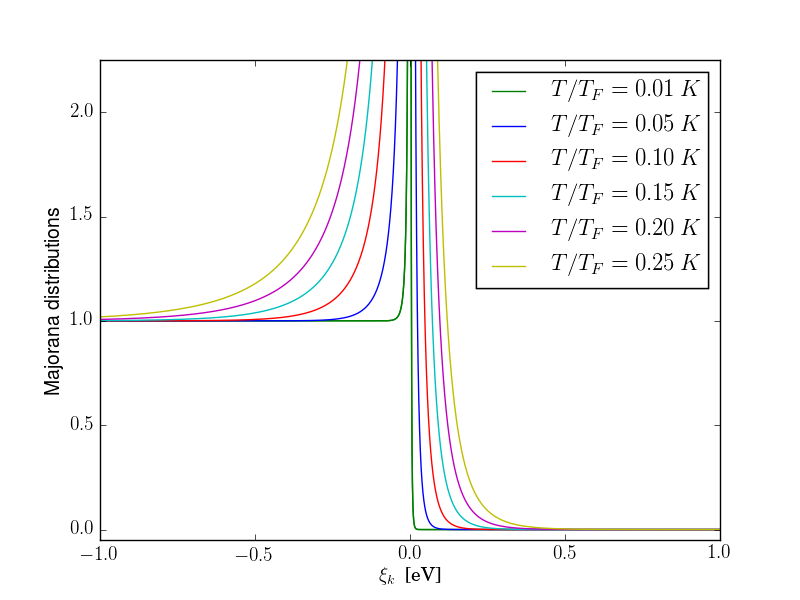}
\caption{}
\label{fig:sub7a}
\end{subfigure}%
\begin{subfigure}{.47\columnwidth}
\centering
\includegraphics[width=1.09\columnwidth]{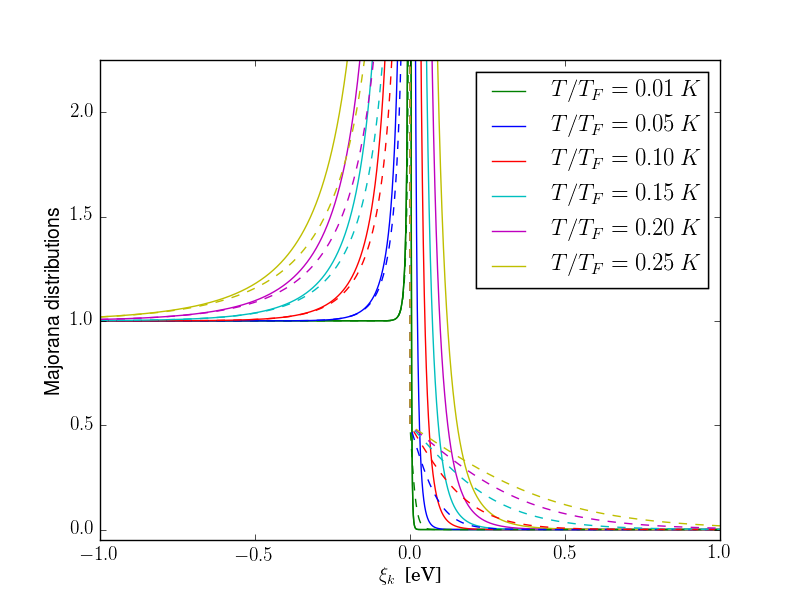}
\caption{}
\label{fig:sub7b}
\end{subfigure}\\[1ex]
\begin{subfigure}{.47\columnwidth}
\centering
\includegraphics[width=1.09\columnwidth]{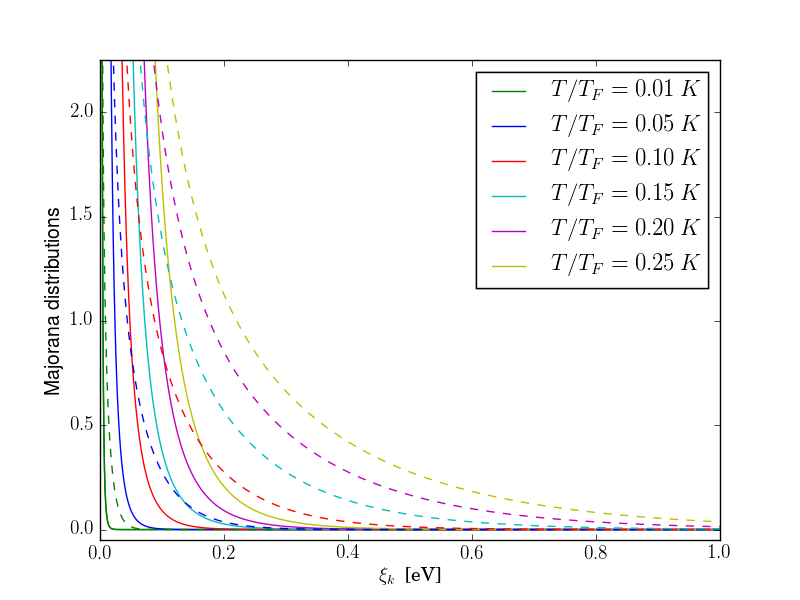}
\caption{}
\label{fig:sub7c}
\end{subfigure}%
\begin{subfigure}{.47\columnwidth}
\centering
\includegraphics[width=1.09\columnwidth]{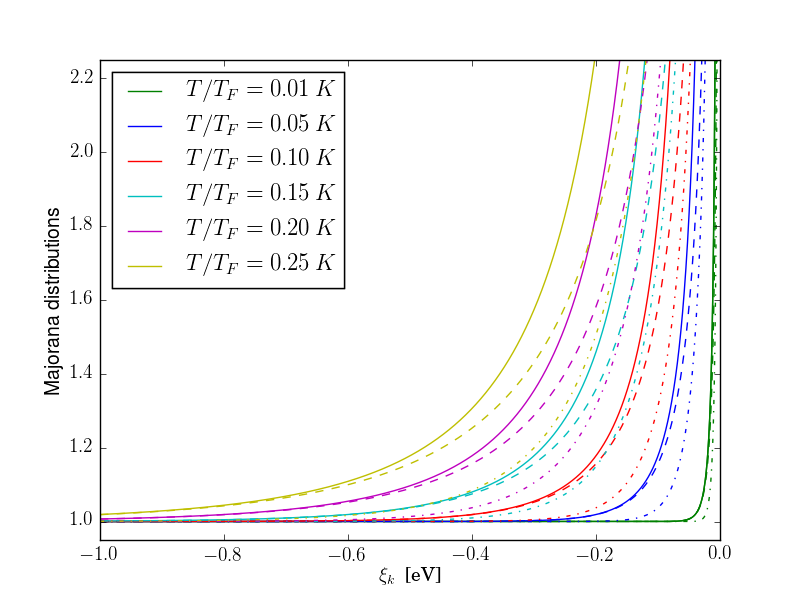}
\caption{}
\label{fig:sub7d}
\end{subfigure}
\caption{{(a) The Majorana\added{-Schwinger} distribution derived from the microscopic second quantization model at various temperatures. (b) The microscopic Majorana\added{-Schwinger} distribution (solid colors) plotted against the Majorana\added{-Schwinger} distribution derived from the statistical approach, with an additional free bosonic contribution below the Fermi surface (dotted colors) for various temperatures. Notice that, in the \added{low$-T$} limit, the statistical and microscopic distributions show similar behavior. The finite\added{$-T$} displacement of the microscopic model's distribution function relative to that from the microscopic model's is the result of a minor bosonic contribution to the Majorana\added{-Schwinger} dispersion. (c) The microscopic Majorana\added{-Schwinger} distribution (solid colors) with a free Fermi-Dirac and Bose-Einstein distribution plotted above the Fermi surface. From such a plot, we can see that the signatures of a sharp Fermi surface are not accounted for by the trivial inclusion of a free boson gas. (d) The Majorana\added{-Schwinger} distribution from the microscopic Hamiltonian (solid colors), the statistical model (dashed colors), and a free Fermi-Dirac and Bose-Einstein distribution (dashed-dotted colors) below the Fermi surface. Better agreement with the microscopic model is seen in the inclusion of our statistical description of the fermionic degrees of freedom than if we consider the presence of a free Fermi gas in the $\xi_k<0$ regime.}}
\end{figure*}


The above is the exact form of the Majorana\added{-Schwinger} distribution, which we see is equivalent to a Fermi-Dirac distribution with an additional bosonic contribution proportional to the average hole occupation. Such bosonic degrees of freedom are the direct result of mutual pairwise annihilation inherent to the non-interacting Majorana\added{-Schwinger} gas. At zero temperature, the Fermi-Dirac distribution $n^{(f)}(\xi_k)$ will go to a Heaviside function, while the quantity 
$n^{(b)}(\xi_k)\left(n^{(b)}(\xi_k)-1\right)$ will go to a constant for $\xi_k<0$ and will go to zero for $\xi_k>0$. Therefore, we see that the Majorana\added{-Schwinger} distribution function is equivalent to the Fermi-Dirac distribution at zero temperature (as we found previously from the statistical model). Low-temperature Fermi-Dirac like behavior in the propagation of stable Bose-Fermi pairs has also been seen in the study of some general Bose-Fermi mixtures \cite{Storozhenko, Daniele}. 

In Fig. \ref{fig:sub7a}, we see a plot of the Majorana\added{-Schwinger} distribution function derived from the microscopic model. A clear difference between this Majorana distribution and the one we derived from building the Boltzmann entropy can be seen in the divergence about the $\xi(k)=0$ point in the former. This is a direct consequence of the emergent bosonic character of the Majorana\added{-Schwinger} system as we raise temperature--i.e., as the \added{MSFs} begin to annihilate. In the statistical model we introduced in the previous sections, we only considered the fermionic degrees of freedom, and therefore excluded the bosonic contribution we see from the microscopic model.

If we wish to compare the Majorana\added{-Schwinger} distribution we derived from our general statistical approach to the one we derived from our modified Kitaev Hamiltonian, we will need to include the emergent bosonic contribution in the former. This can be done by adding a bosonic term to the Majorana\added{-Schwinger} distribution we derived from our statistical argument:

\begin{align}
n_{k}^{(m)}&=\bigg\{\bigg(1+n^{(b)}_k(|\xi_k|)\bigg)\Theta(-\xi_k)+\left(n^{(f)}_k(\xi_k)\right)\Theta(\xi_k)\bigg\}
\end{align}

\noindent  We take the absolute value in the exponential above because the bosonic contribution is measured with respect to $\mu$ below the Fermi surface. In Fig. \ref{fig:sub7b}, we see the two Majorana\added{-Schwinger} distribution functions calculated for a small range of temperatures, where we see agreement between the statistical and microscopic derivations of the distributions in the low temperature limit about the $\xi_k=0$ regime. Note that the Majorana\added{-Schwinger} distribution we derived previously captures the features of a sharp Fermi surface in the microscopic model if we exclude the bosonic contributions $n^{(m)}(\xi_k)>1$. Such a sharp Fermi surface is the direct result of mutual pairwise annihilation in the fermionic system, and is not the result of the emergent free boson gas. We can see this in \ref{fig:sub7c}, where we have plotted the microscopic model (solid color) against a free fermion and free boson distribution for $\xi_k>0$. We can easily see that the full bosonic contribution does not contribute to the sharp finite-temperature Fermi surface which is the hallmark of the Majorana\added{-Schwinger} gas. Such a sharp Fermi level is, however, included in our statistical theory. Note that the inclusion of a fully-filled Fermi sea below $\xi_k=0$ (as predicted by the statistical model) is in better agreement with the microscopic distribution than the inclusion of a free Fermi sea with finite-temperature smearing. This can be seen in Fig. \ref{fig:sub7d}. Also note that the apparent displacement of the microscopic model's "Fermi surface" from the statistical model's is a direct result of the small emergent bosonic contribution not captured in the statistical analysis. In the next section, we will directly compare the temperature dependence of the chemical potential in both models and see that they agree with reasonable precision. }

\section{IV. Thermodynamics of the free Majorana\added{-Schwinger} gas}

\subsection{Thermodynamic observables from the statistical model of a \added{number-conserving} Majorana gas}


With the Majorana\added{-Schwinger} distribution function derived, we can now turn to the thermodynamics of the non-interacting Majorana\added{-Schwinger} gas. First, note that, from the zero-temperature behavior of the Majorana\added{-Schwinger} distribution function discussed above, the relation between the total particle number and the Fermi energy is identical to the Fermi case. Hence, the Fermi energy of the Majorana\added{-Schwinger} gas at zero temperature is identical to that of the Fermi gas. { Also note that we calculate all thermodynamic quantities at a fixed temperature $T$; i.e., in the absence of a temperature gradient. As discussed in the previous section, this allows us to preserve a $U(1)$ symmetry in the low-temperature Majorana\added{-Schwinger} system and hence a well-defined particle number. Finally, be aware that in this section we are only considering the fermionic degrees of freedom of the Majorana\added{-Schwinger} system, which we studied exclusively in Section III A of this article via a modified fermion combinatorics. As we will see later on, the bosonic contribution is negligible up to linear order in temperature.}

As we progress to non-zero temperature, the thermodynamics of the \added{Majorana-Schwinger gas} differs from that of the Fermi gas due to the sharply-defined Fermi surface at finite-temperature we found in the Part III of this paper. As such, we have to consider the regions $\epsilon_{p\sigma}<\mu$ and $\epsilon_{p\sigma}>\mu$ separately in the calculation of the total particle number:

\begin{align}
N&=\int_0^\mu n_{p\sigma}(T\not=0)g(\epsilon)d\epsilon+\int_\mu^\infty n_{p\sigma}(T\not=0)g(\epsilon)d\epsilon\notag\\
&\sim V \left(\frac{2}{3}\mu^{3/2}+\Gamma(3/2)F_{3/2}(\mu/T,\,\mu)\right)\label{30}
\end{align}
where we have taken $\epsilon_{p\sigma}\equiv \epsilon$ for simplicity and utilized the incomplete Fermi-Dirac function:

\begin{align}
F_{\gamma+1}(\mu/T,\,\mu)=\frac{1}{\Gamma(\gamma+1)}\int_\mu^\infty \frac{\epsilon^\gamma}{\exp\left((\epsilon-\mu)/T\right)+1}d\epsilon
\end{align}

\noindent The incomplete Fermi-Dirac function is evaluated for general parameters in Appendix C. The result is an infinite sum of complete Fermi-Dirac functions with a fugacity of one:
\begin{align}
F_{\gamma+1}(\mu/T,\,\mu)&=\sum_{k=0}^\infty \frac{1}{(\gamma-k)!}T^{k+1}\mu^{\gamma-k}F_{k+1}(0,\,0)\label{21}
\end{align}
From the above, we can easily see that, in the low-temperature limit,

\begin{table*}
\caption{\label{tab:table1} Observables in the non-interacting, non-relativistic 3D, 2D, and 1D Majorana\added{-Schwinger (MS)} and Fermi\added{-Dirac (FD)} gases. Note that the energy and specific heat for the Majorana\added{-Schwinger} system is nearly identical to that of the Fermi system in all dimensions. However, the chemical potential and entropy in the Majorana\added{-Schwinger} gas differ greatly from the fermionic. \added{The former} harbors an extra term that is linear in temperature. This term subsequently leads to a residual entropy of $\log 2$ per particle that is not found in the Fermi gas.}
\begin{tabular}{ l  c  r }
\hline\hline\\
Observable & Non-relativistic 3D MS gas & Non-relativistic 3D FD gas\\
\hline\\
$\mu/\epsilon_F$ & $\approx
1-\frac{T}{T_F}\log 2 -\frac{\pi^2}{12}\left(
\frac{1}{2}-\frac{6}{\pi^2}(\log 2)^2\right)\frac{T^2}{T_F^2}
$ & 
$\approx 
1-\frac{\pi^2}{12}\frac{T^2}{T_F^2}
$\\\\
$U/U_0$ & $\approx 
1+\frac{5\pi^2}{12}\left(\frac{1}{2}-\frac{3}{2\pi^2}(\log 2)^2\right)\frac{T^2}{T_F^2}$ & 
$\approx 1+\frac{5\pi^2}{12}\frac{T^2}{T_F^2}$\\\\
$C_v/N$ & 
$\approx \frac{\pi^2}{2}\left(
\frac{1}{2}-\frac{3}{2\pi^2}(\log 2)^2
\right)\frac{T}{T_F}$ & $\approx \frac{\pi^2}{2} \frac{T}{T_F}$\\\\
$S/N$ & $\approx \frac{\pi^2}{2}\left(\frac{1}{2}-\frac{9}{4\pi^2}(\log 2)^2\right)\frac{T}{T_F}+\log 2 $ & $\approx \frac{\pi^2}{2}\frac{T}{T_F}$ \\\\
\hline \hline \\
Observable & Non-relativistic 2D MS gas & Non-relativistic 2D FD gas \\
\hline \\
$\mu/\epsilon_F$ & $=1-\frac{T}{T_F}\log 2$ & $=T \log \left(\exp(T_F/T)-1\right)$\\\\
$U/U_0$ & $=1+\frac{\pi^2}{3}\left(\frac{1}{2}-\frac{3}{\pi^2}(\log 2)^2\right)\frac{T^2}{T_F^2} $& $\approx 1+\frac{\pi^2}{3}\frac{T^2}{T_F^2}$\\\\
$C_v/N$ & $=\frac{\pi^2}{3}\left(\frac{1}{2}-\frac{3}{\pi^2}(\log 2)^2 \right) \frac{T}{T_F}$ & $\approx \frac{\pi^2}{3}\frac{T}{T_F}$\\\\
$S/N$ & $=\frac{\pi^2}{3}\left(\frac{1}{2}-\frac{3}{\pi^2}(\log 2)^2\right)\frac{T}{T_F}+\log 2$ & $ \approx \frac{\pi^2}{3}\frac{T}{T_F}$\\\\
\hline \hline \\
Observable & Non-relativistic 1D MS gas & Non-relativistic 1D FD gas \\
\hline \\
$\mu/\epsilon_F$ & $\approx 1-\frac{T}{T_F}\log 2 +\frac{\pi^2}{12}\left(\frac{1}{2}-\frac{6}{\pi^2}(\log 2)^2\right)\frac{T^2}{T_F^2}$ & $ \approx 1+\frac{\pi^2}{12}\frac{T^2}{T_F^2}$ \\\\
$U/U_0$ & $\approx 1+\frac{\pi^2}{4}\left(\frac{1}{2}-\frac{6}{\pi^2}(\log 2)^2\right)\frac{T^2}{T_F^2}$ & $\approx 1+\frac{\pi^2}{4}\frac{T^2}{T_F^2}$\\\\
$C_v/N$ & $\approx \frac{\pi^2}{6}\left(\frac{1}{2}-\frac{6}{\pi^2}(\log 2)^2\right)\frac{T}{T_F}$ & $\approx \frac{\pi^2}{6}\frac{T}{T_F}$\\\\
$S/N$ & $\approx \frac{\pi^2}{6}\left( \frac{1}{2}-\frac{6}{\pi^2}(\log 2)^2\right)\frac{T}{T_F}+\log 2$ & $\approx \frac{\pi^2}{6}\frac{T}{T_F}$\\\\
\hline \hline
\end{tabular}
\end{table*}

\begin{table*}
\caption{\label{tab:table2}  Observables in the non-interacting, ultra-relativistic 3D, 2D, and 1D Majorana\added{-Schwinger (MS)} and Fermi\added{-Dirac (FD)} gases. We see that the ultra-relativistic system behaves similarly to the non-relativistic gas, with the Majorana\added{-Schwinger} internal energy and specific heat identical to the Fermi system except with the temperature quadratic in temperature modified by a correction factor. The chemical potential retains a $-\frac{T}{T_F}\log 2$ term seen in the non-relativistic system, and thus the entropy has a residual $\log 2$ term.}
\begin{tabular}{ l  c  r }
\hline \hline\\
Observable & Ultra-relativistic 3D MS gas & Ultra-relativistic 3D FD gas\\
\hline\\
$\mu/\epsilon_F$ & $\approx 1-\frac{T}{T_F}\log 2 -\frac{\pi^2}{3}\left(\frac{1}{2}-\frac{6}{\pi^2}(\log 2)^2\right)\frac{T^2}{T_F^2}$ & 
$\approx 1-\frac{\pi^2}{3}\frac{T^2}{T_F^2}$\\\\
$U/U_0$ & $\approx 1+2\pi^2 \left(\frac{1}{6}+\frac{1}{\pi^2}(\log 2)^2\right)\frac{T^2}{T_F^2}$ & 
$\approx 1+2\pi^2 \frac{T^2}{T_F^2}$\\\\
$C_v/N$ & 
$\approx 3\pi^2 \left(\frac{1}{6}+\frac{1}{\pi^2}(\log 2)^2\right)\frac{T}{T_F}$ & $\approx 3\pi^2 \frac{T}{T_F}$\\\\
$S/N$& $\approx \frac{\pi^2}{2}\frac{T}{T_F}+\log 2$ & $\approx \frac{7\pi^2}{3}\frac{T}{T_F}$ \\\\
\hline \hline \\
Observable & Ultra-relativistic 2D MS gas & Ultra-relativistic 2D FD gas \\
\hline\\
$\mu/\epsilon_F$ & $\approx 1-\frac{T}{T_F}\log 2 -\frac{\pi^2}{6}\left(\frac{1}{2}-\frac{6}{\pi^2}(\log 2)^2\right)\frac{T^2}{T_F^2}$ & 
$\approx 1-\frac{\pi^2}{6}\frac{T^2}{T_F^2}$\\\\
$U/U_0$ & $\approx 1+\frac{\pi^2}{4}\frac{T^2}{T_F^2}$ & 
$\approx 1+\pi^2 \frac{T^2}{T_F^2}$\\\\
$C_v/N$ & 
$\approx \frac{\pi^2}{3}\frac{T^2}{T_F^2}$ & $\approx \frac{4\pi^2}{3}\frac{T^2}{T_F^2}$\\\\
$S/N$ & $\approx \frac{7\pi^2}{6}\left(\frac{2}{7}-\frac{6}{7\pi^2}\left(\log 2\right)^2\right)\frac{T}{T_F}+\log 2$ & $\approx \frac{7\pi^2}{6}\frac{T}{T_F}$ \\\\
\hline \hline \\
Observable & Ultra-relativistic 1D MS gas & Ultra-relativistic 1D FD gas \\
\hline \\
$\mu/\epsilon_F$ & $=1-\frac{T}{T_F}\log 2$ & $=T \log \left(\exp(T_F/T)-1\right)$\\\\
$U/U_0$ & $=1+\frac{\pi^2}{3}\left(\frac{1}{2}-\frac{3}{\pi^2}(\log 2)^2\right)\frac{T^2}{T_F^2} $& $\approx 1+\frac{\pi^2}{3}\frac{T^2}{T_F^2}$\\\\
$C_v/N$ & $=\frac{\pi^2}{3}\left(\frac{1}{2}-\frac{3}{\pi^2}(\log 2)^2 \right) \frac{T}{T_F}$ & $\approx \frac{\pi^2}{3}\frac{T}{T_F}$\\\\
$S/N$ & $=\frac{\pi^2}{3}\left(\frac{1}{2}-\frac{3}{\pi^2}(\log 2)^2\right)\frac{T}{T_F}+\log 2$ & $ \approx \frac{\pi^2}{3}\frac{T}{T_F}$\\\\
\hline \hline
\end{tabular}
\end{table*}

\begin{align}
&F_{3/2}(\mu/T,\,\mu)\notag\\
&\approx \frac{1}{(1/2)!}T\mu^{1/2}F_1(0,\,0)+\frac{1}{(1/2-1)!}T^2 \mu^{1/2-1}F_2(0,\,0)\notag\\
&=\frac{2\log 2}{\sqrt{\pi}}\mu^{1/2}T +\frac{\pi^{3/2}}{12}\frac{T^2}{\mu^{1/2}}
\end{align}
Recalling the form of Eqn. \eqref{30}, we find the relation

\begin{align}
\frac{2}{3}\epsilon_F^{3/2}\approx\frac{2}{3}\mu^{3/2}+T\mu^{1/2} \log 2 +\frac{\pi^2}{24}\frac{T^2}{\mu^{1/2}}
\end{align}
This might appear counterintuitive, because the Majorana\added{-Schwinger} gas does not conserve particle number due to particle-particle annihilation. We can get around this issue by assuming that the Majorana\added{-Schwinger} system is in chemical equilibrium with an external particle reservoir and restricting ourselves to the low-temperature regime. We thus have the ability to describe a system with a constant mean particle number that still exhibits the Majorana mutual annihilation and, as such, a deviation from Fermi-Dirac statistics.

The main thermodynamic observables of the non-relativistic Majorana\added{-Schwinger} gas is shown in Table \ref{tab:table1} side-by-side with the Fermi gas observables. The derivation of these quantities is given in Appendix D. For the sake of completeness, we also include the thermodynamic observables of the ultra-relativistic Majorana\added{-Schwinger} gas in Table \ref{tab:table2}.

When we compare the results of the two systems, we notice that the majority of the terms quadratic in temperature are nearly identical to the corresponding terms in the Fermi gas, except that in the former they are reduced by a factor less than one half. From the results in one, two, and three dimensions, we can suggest the following form of the $d$-dimensional Majorana\added{-Schwinger} correction factor:
\begin{align}
\gamma_d=\frac{1}{2}-\frac{3}{2^d}\left(\frac{2}{\pi} \log 2\right)^2\label{25}
\end{align}
All thermodynamic quantities are reduced by the same factor in the 1D case. In the 2D system, the quadratic temperature dependence in the chemical potential disappears (as it does in the 2D Fermi gas), while the correction factors in the 3D chemical potential and entropy differ slightly from the term in the internal energy and chemical potential. These discrepancies are more than likely the result of the repeated approximations and series expansions used in the 3D system as opposed to the simpler 2D or 1D systems. 

The most shocking difference between the Majorana\added{-Schwinger} and Fermi\added{-Dirac} gases is the linear dependence in temperature seen in the former's chemical potential. Such a chemical potential results in a constant $\log 2$ term in the entropy per particle. Even more interesting is that this term appears in the same form in all dimensions, and is thus a fundamental signature of the Majorana\added{-Schwinger} gas. Such a residual term in the entropy is the result of a two-fold degeneracy in the occupation of each Majorana\added{-Schwinger} ground state; e.g., unlike the non-interacting Fermi system, any microstate has a finite probability of being both occupied or unoccupied. This residual entropy is similar to that seen in water ice \cite{Pauling} or quantum spin ice in magnetic pyrochlore materials \cite{Gingras} except for the fact that, in this system, the zero-point degeneracy is not the result of geometric frustration, but is instead caused by mutual particle-particle annihilation. From this residual entropy, we conclude that the Majorana\added{-Schwinger} gas in the limit of zero external temperature behaves identically to that of a two-level system, with the degeneracy the result of the interplay between Pauli correlation and the particle-particle annihilation. When the population of Majorana\added{-Schwinger} fermions at the higher energy state (i.e., either separated or annihilated) is greater than that at the lower energy, the system will experience a negative internal temperature. As a result, the Majorana system in this limit is highly unstable and might be considered out of equilibrium. { It is also worth noting that a zero-point thermodynamic entropy is also seen in the neutral Majorana sea predicted in the Kondo insulator SmB$_6$ \cite{Coleman, Kivelson}}.

From the above analysis, we can now see clear differences between the non-interacting Majorana\added{-Schwinger} and Fermi gases. At zero temperature, the Majorana\added{-Schwinger} gas behaves as a Fermi gas with a residual entropy caused by the interplay of particle-particle annihilation and fermionic Pauli correlation. 
 The system's stability depends on the relative energy-cost of annihilation, i.e. if if the particles prefer to annihilate each other or remain in distinct energy states. The system therefore has two temperature scales: one coming from the "frustration" of the system and one coming from the regular thermodynamic energy. As we raise the external temperature, the system behaves similar to that of a Fermi\added{-Dirac} system, although now in a slightly-modified form to account for the particle-particle annihilation. This annihilation is most apparent in the chemical potential, which experiences a universal term that goes linearly with temperature and is independent of dimension. Particle annihilation also comes into play in the internal energy and specific heat, which experiences a decline in terms quadratic in temperature on the order of the correction term $\gamma_d$. The correction term decreases with decreasing dimension, which illustrates that the thermodynamics of the Majorana\added{-Schwinger} gas is dominated by particle-particle annihilation as we decrease dimensionality.

We can check for consistency by seeing if the derivative of the free energy $F=U-TS$ with respect to the particle number is the chemical potential. Using the expressions above, we see that the 2D Majorana\added{-Schwinger} free energy is given by
\begin{align}
F&=\frac{N^2\pi \hbar^2}{Am}\left\{
1-\left(
(\log 2)^2 -\frac{\pi^2}{6}
\right)\frac{T^2}{T_F^2}
\right\}\notag\\
&\phantom{=}-N\left(\frac{\pi^2}{6}-(\log 2)^2\right)\frac{T^2}{T_F}-NT \log 2
\end{align}
As such, the derivative of the above with respect to $N$ yields the following relation, which agrees with Eqn. \eqref{23}:
\begin{align}
\frac{\partial F}{\partial N}\bigg|_{T,\,V}&=\epsilon_F \left(1-\frac{T}{T_F}\log 2\right)
\end{align}

{ \subsection{
Comparison of the microscopic and statistical models' chemical potentials
}

In the previous section, we saw that one of the most noticeable differences between the free Majorana\added{-Schwinger} gas and the free Fermi gas is the presence of a linear-temperature dependence in the chemical potential of the former. To first order, the chemical potential of the Majorana\added{-Schwinger} gas goes as $\mu=\epsilon_F-T\log 2\approx \epsilon_F- 0.693 T$ for all dimensions. To confirm this result, let us compare this value to the temperature-dependence of the chemical potential from the microscopic model of the Majorana\added{-Schwinger} gas we derived in a previous section. Recall that the free Majorana\added{-Schwinger} distribution $n^{(m)}(\xi_k)$ can be written in terms of fermionic and bosonic distributions (see Eqn. \eqref{micro}). If we consider only the fermionic degrees of freedom in the Majorana\added{-Schwinger} distribution function, the chemical potential should mark the point where $n^{(m)}(\xi_k)=1/2$ in the finite temperature limit. Solving for the value of $\xi_k=\epsilon_k-\mu$ where the distribution equals one half occupation, we find the following:
\begin{align}
&\left(\epsilon_k-\mu\right)\bigg|_{n^{(m)}(\xi_k)=1/2}\notag\\
&=T \log \left(\frac{1}{3} \left(1+\sqrt[3]{19-3 \sqrt{33}}+\sqrt[3]{19+3
   \sqrt{33}}\right)\right)\notag\\
   &\approx 0.6093 T
\end{align}
Note that the dispersion of the total Majorana\added{-Schwinger} system in the microscopic model contains two components: the fermionic contribution $\epsilon^{(f)}\approx \epsilon_F$ and the bosonic contribution from mutual annihilation $\epsilon^{(b)}$. The chemical potential is the same for the fermionic and bosonic contributions, as we consider them to be in thermal equilibrium with each other \cite{Herrmann}. We therefore see that the microscopic model yields the following temperature dependence for the chemical potential:
\begin{align}
\epsilon_F+\epsilon^{(b)}-\mu\approx 0.6093 T
\end{align}
Compare this with the result we found in the statistical model, which does not consider the bosonic degrees of freedom in the Majorana\added{-Schwinger} system; namely, $\epsilon_F-\mu\approx 0.693 T$. Both the microscopic and statistical models show similar temperature dependence up to linear order, and both models have near-identical multiplicative pre-factors. This tells us that the fermionic degrees of freedom are dominant in the Majorana\added{-Schwinger} gas above the Fermi level, and illustrates further agreement between our statistical approach and the emergent behavior of our microscopic toy Hamiltonian.
}

\section{V. Agreement with present theories and the possibility of experimental realization}

\subsection{Signatures of Majorana\added{-Schwinger} statistics in condensed matter systems}

\added{In this work, we have so far introduced a new formalism to describe a non-interacting, many-body system of Majorana fermions. Whereas the anyonic or SYK model describes a purely fermionic realization of self-conjugate, spin-$1/2$ particles, the Majorana-Schwinger formalism extends the fermionic Hilbert space to include the bosonic contribution from emergent annihilation while simultaneously allowing us to consider a well-defined statistical description in the non-interacting limit. This has led us to a closed form of the Majorana-Schwinger distribution function, and emergent macroscopic thermodynamics that differs from that of complex fermions. However, the question remains if the MZMs studied in certain condensed matter systems can be modeled as a free Majorana-Schwinger gas. { Whereas the conventional SYK model requires strong interactions and zero chemical potential\cite{Maldacena,Kitaev_SYK,Pikulin}}, it has already been noted that the MZMs in topological superfluids might be well-defined in a number-conserving theory if we include the effects from the condensate fraction in the usual mean-field description\cite{Leggett_Lin1,Leggett_Lin2}. By making an analogy with momentum-conservation in the high-energy limit, we have taken an alternative direction in the study of number-conserving Majorana fermions by defining an initially fermionic system that experiences mutual annihilation only in the finite-temperature limit, and which does not necessarily require the presence of topological order. Below we consider several possible materials that might support or show signatures of Majorana-Schwinger statistics.}

\subsubsection{\added{Superconducting systems}}
 
\added{Due to their particle-hole symmetry,} Bogoliubov quasiparticles are a natural candidate for \added{a condensed matter analog of} Majorana fermions \added{that obey the spin-statistics theorem}. The \added{Majorana-like} nature of these particles may be verified via a correlation of two electron beams after repeated Andreev reflection with a superconducting contact, which imposes a particle-hole symmetry and \added{subsequent} pairwise annihilation \added{without the need for topologically non-trivial correlation} \cite{Beenakker2}. Similar studies have been suggested with single electron and hole propagation in a quantum Hall edge state, so as to achieve a zero-frequency noise measurement and, thus, more reliable data \cite{Ferraro}. 
Measurements of thermodynamic quantities, such as the internal energy, the momentum profile, or the fugacity, may be easily explored in a gas of non-interacting Bogoliubov quasiparticles in much the same manner as they are found in an ultracold Fermi gas \cite{DeMarco1, DeMarco2}, \added{giving clear indication if these particles obey MS statistics}. Similar thermodynamic measurements might be used to prove the existence of a gas of Majorana\added{-Schwinger fermions} in Dirac-type s-wave induced topological superconductors \cite{Chamon}.

\subsubsection{\added{Topological matter}}

Signatures of Majorana\added{-Schwinger} thermodynamics might also be found in a many-body system of Majorana zero modes. Although we have explicitly shown that non-Abelian statistics differs greatly from that of our Majorana\added{-Schwinger} system, the simplest example of an Abelian Majorana fermion exhibits both a nontrivial statistical phase and charge conjugation \cite{Freedman}. The former will yield Haldane-Wu "intermediate" statistics in the absence of the latter. A Majorana\added{-Schwinger like} mutual annihilation condition on the particles imposes a modulo-2 occupation of the microstates, and thus the Haldane-Wu statistics reduces to the Majorana\added{-Schwinger} statistics derived above. One might argue that there will no longer be a fermionic ground state due to a repressed Pauli correlation in the anyonic system, but this disagrees with current AFM images of Majorana bound states in Fe chains on a superconducting Pb surface \cite{Pawlak}. The AFM map shows direct evidence of a repulsive Pauli effect in the vicinity of the \added{MZMs}, and thus appears to support the argument for a fermionic ground state in the thermodynamics of the Majorana zero modes \added{and a possible realization of Majorana-Schwinger statistics}.

Application of \added{Majorana-Schwinger} thermodynamics to MZMs is \added{also} supported by recent research from Morais Smith et. al. on the Hill thermodynamics of the 1D Kitaev chain, the 2D Kane-Mele (KM) model, and the 3D Bernevig-Hughes-Zhang (BHZ) model in the topological regime \cite{Kempkes1, Kempkes2, Kempkes3}. Hill thermodynamics divides the thermodynamic potential of a finite-size many-body system into a potential for an infinite system and a separate sub-division potential containing finite-size effects \cite{Hill}, the former of which describes the bulk behavior and the boundary described by the latter. The topological regimes of these models host bound and unbound Majorana edge modes, and hence the thermodynamics of their boundaries should agree with our model. Indeed, in all three materials, Hill thermodynamics yields observables that are strikingly similar to Majorana\added{-Schwinger} statistics at low temperature. The specific heat of the KM and BHZ edge states have a linear temperature dependence (as seen in our model), and the low temperature behavior of the BHZ specific heat $C_v/k_BT$ in the topological phase goes as $\pi/3\approx 1.05$, which is fairly close to the 2D Majorana\added{-Schwinger} correction factor $\frac{\pi^2}{3}\left(\frac{1}{2}-\frac{3}{\pi^2}(\log 2)^2\right)\approx 1.16$. Perhaps the most notable aspect of the Smith group's study is the entropy of the Kitaev chain boundary, which starts at a value of $\log 2$ in the topological phase and then decreases to zero with increasing temperature. Using Eq. \eqref{25}, we see that setting $d=0$ (corresponding to the boundary states) yields a value of $\gamma_0=\frac{1}{2}-\frac{3}{2^0}\left(\frac{2}{\pi}\log 2\right)^2\approx -0.084$. The negative $d=0$ Majorana correction factor and $\log 2$ residual entropy agree with the results from Hill thermodynamics. Note that the study of Smith et. al. only considers an Abelian Berry phase, and does not consider a non-Abelian exchange statistics (\added{similar to our Majorana-Schwinger model}). 

\added{Beyond the idealistic Kitaev chain model and its higher-dimensional extensions, further experimental evidence might come from the many-body effects of Majorana edge states in topological superconductors, such as FeTe$_{1-x}$Se$_x$ \cite{Ding1, Ding2} and Cu$_x$Bi$_2$Se$_3$ \cite{Sasaki,Levy}, which are defined by the presence of MZMs and could lead to the realization of Majorana statistics beyond the Dirac-type electron field amplitudes of Chamon et. al. \cite{Chamon}. }

\subsubsection{\added{Kondo materials}}

In a similar vein, a recent study of the dielectric state of a superconductor under topological failure of superflow shows evidence of \added{number-conserving} Majorana\added{-Schwinger} thermodynamics, with the linear specific heat of the Kondo insulator samarium hexaboride (SmB$_6$) theorized to be the result of a neutral Majorana Fermi sea \cite{Erten}. { A neutral Majorana sea in SmB$_6$ was first proposed by Emery and Kivelson in the context of a two-channel Kondo model, which was then expanded upon by Coleman et. al. \cite{Kivelson, Coleman}.} Such a \added{self-conjugate representation of} the Kondo insulator is in stark contrast to the regular \added{complex} fermion phase, which is strongly interacting and may only be described be a weakly interacting Fermi gas under a non-trivial unitary transformation \cite{Eder, Ostlund}. Experimental studies of SmB$_6$ likewise show a surprisingly low effective mass $m/m_e$ of $0.119\pm 0.007$, $0.129\pm 0.004$, and $0.192\pm 0.005$ for three different values of the oscillatory magnetic torque's frequency, given as $\sim 35$ T, $300$ T, and $450$ T, respectively \cite{Li}. Comparing this with the effective mass $m/m_e=0.225\pm 0.011$ for the electron pocket of the semimetallic compound EuB$_6$ \cite{Aronson} (which are in good agreement with band-structure calculations), we see the effective masses of SmB$_6$ and EuB$_6$ differ by a factor $\delta$ of approximately $0.529\pm 0.013$, $0.573\pm 0.012$, and $0.853\pm 0.012$ for the three different magnetometric frequencies (in increasing order). Erten et. al. have suggested that this small effective mass is due to the fact that the Majorana sea chiefly originates from the conduction electron band \cite{Erten}. \added{By extending the Majorana Hilbert space to preserve total number conservation while simultaneously permitting self-conjugation, our} Majorana\added{-Schwinger} statistics similarly predicts a low effective mass due to the form of the 3D correction factor to the specific heat. This factor, given by $\gamma_3=\frac{1}{2}-\frac{3}{2\pi^2}(\log 2)^2\approx 0.427$, is on the same scale as the experimental factor $\delta$ for SmB$_6$, and might be an indicator of dominant free Majorana\added{-Schwinger} gas behavior at lower oscillatory magnetic torque frequencies. { Further evidence of Majorana\added{-Schwinger} thermodynamics in SmB$_6$ can be seen by probing the temperature dependence of quantum oscillation amplitudes in the bulk. \footnote{We thank Suchitra Sebastian for bringing this study to our attention} Tan et. al. have shown that there exists a residual density of states at the Fermi energy in this material, and that quantum oscillations increase rapidly for decreasing temperature, as opposed to a saturation of oscillations as predicted by Fermi-Dirac statistics \cite{Lifshitz, Shoenberg, Tan}. In a system whose fundamental excitations are described by Fermi-Dirac statistics, oscillations will gradually decrease as we raise temperature due to a "smearing-out" of the Fermi-Dirac distribution \cite{Shoenberg}. The sharply defined finite-temperature Fermi surface in the Majorana Fermi sea present in SmB$_6$ may then be a possible explanation for the amplification of quantum oscillations in this material.} 

\subsubsection{\added{Kitaev honeycomb lattices}}

Another promising candidate for the Majorana\added{-Schwinger} many-body system might be found in the fractionalized excitations of a Kitaev honeycomb lattice \cite{Kitaev_honey,Vala,Ville}. Such a system consists of strongly interacting spin$-1/2$ fermions that can be exactly mapped to \added{a free Majorana representation}. Inelastic neutron scattering and Raman spectroscopy have yielded firm evidence for fractionalized Majorana excitations in the spin lattice $\alpha$-RuCl$_3$ \cite{Do} and the iridates $\beta$- and $\gamma$- Li$_2$IrO$_3$ \cite{Glamazda}, so it \added{seems promising to expect Majorana-Schwinger} thermodynamics to characterize these systems. \added{The Kitaev honeycomb lattice} exhibits two separate ground states: a gapped phase harboring Abelian anyons and a gapless spin liquid phase, the latter of which being able to gap out into a topological spin liquid hosting non-Abelian anyons in the presence of a magnetic field.  Generalizations of the Kitaev model on the three-dimensional hyperoctagon lattice have shown that the gapless spin liquid phase hosts a two-dimensional Fermi surface of itinerant Majorana fermions \cite{Trebst, Trebst2}. The presence of such a "Majorana metal" of spinons agrees with the thermodynamics of our \added{number-conserving} theory. On the computational side, quantum Monte Carlo simulations of a Kitaev honeycomb model show a linear temperature dependence in the specific heat at the crossover between itinerant and localized Majorana particles, in stark contrast to the predicted quadratic behavior from the Dirac semimetallic dispersion \cite{Nasu}. A $T^2$ behavior is only apparent in the low-temperature region, which is dominated by thermal fluctuations of fluxes of localized Majorana fermions. The linear-$T$ specific heat could be the result of a dominant Majorana\added{-Schwinger} gas behavior in the itinerant Majorana fermions of the Kitaev model. Moreover, experiments in Raman spectroscopy on $\alpha$-RuCl$_3$ have yielded possible evidence of a Fermi-like Majorana distribution function through a measurement of the magnetic contribution to the phonon linewidth \cite{Burch}. A recent study on $\alpha$-RuCl$_3$ has already shown striking agreement with our present theory, where the sum of loss and gain intensity in the XY Stokes and anti-Stokes spectra hints at low-temperature fermionic behavior of the finite-temperature Majorana particles, with increasing bosonic behavior as the temperature is raised \cite{Yiping}. The effective excitations of a Kitaev honeycomb model in the finite temperature limit therefore holds the potential to be an exciting and fruitful realization of our Majorana\added{-Schwinger} gas model.
\\\\
\added{All of the above materials are unique in the sense that, even though they harbor low-temperature excitations that respect the Majorana reality condition, their statistical behavior appears to contradict the predictions of the anyonic model; i.e., that a $U(1)$ symmetry is broken as soon as self-conjugacy appears. Hill thermodynamic analysis of MZMs in the Kitaev chain, experimental signatures of quantum oscillations in samarium hexaboride, and Raman spectroscopy measurements of the finite-temperature behavior of itinerant Majorana fermions in $\alpha$-RuCl$_3$ all appear to support a fermionic ground state and fermionic-like thermodynamics despite supporting Majorana fermions. Our theory shows a way to realize self-conjugacy while retaining total particle number conservation and hence, in the low-temperature limit, an explanation for this complex fermion-like behavior.}
%


\subsection{Detection of Majorana thermodynamics via supernovae neutrino emission and the cosmic neutrino background}

Whereas there is strong evidence (both theoretical and experimental) for Majorana particles in superconducting and topological matter, the existence of a Majorana fermion in the Standard Model is debatable. As mentioned in the introduction, the most promising candidate for a fundamental Majorana fermion is the neutrino, which is postulated to have a right-handed Majorana mass on the order of the GUT scale (and, subsequently, a very small mass for the left-handed species by the seesaw mechanism) \cite{Mohapatra,Tsutomu,Lavoura}. The experimental detection of the Majorana-like nature of neutrinos is, however, exceedingly difficult, with the verification of neutrinoless double-$\beta$ decay remaining inconclusive \cite{Agostini, GERDA}. The Majorana\added{-Schwinger} thermodynamics described above offers a possibly simpler verification of the Majorana behavior of neutrinos, \added{as the MSF model (unlike the case of the MZM or SYK) mirrors total momentum conservation. If neutrinos are truly Majorana fermions, their mutual annihilation at some finite temperature should lead to a pair of photons, as neutrinos have been shown to have some finite mass\cite{Fukuda}.} The detection of thermodynamic or statistical signatures in a \added{low-temperature} neutrino gas could tell us whether or not their behavior is inconsistent with the regular Fermi gas. The main problem with this methodology is that we require the neutrino system to be dense enough to allow for \added{the} degeneracy of the quantum particles to become significant, and we are faced with a limited number of neutrino systems that could harbor Majorana\added{-Schwinger} statistics. \added{Nevertheless, we outline some possible ways to realize Majorana fermion detection in an astrophysical context with the predictions of our Majorana-Schwinger statistics. }

\subsubsection{\added{Supernovae}}
One of the most likely sources of a dense gas of neutrinos is from a type-II supernovae, where the shock wave formed between the collapsing interior and the outer layers of the star's iron core breaches the stellar neutrino layers and produces a large outburst of electron neutrinos \cite{Janka1, Zuber}. The best evidence for this neutrino emission is from the supernova of a blue B3I supergiant on Feb 23 1987 in the Large Magellanic Cloud (dubbed SN1987A), which was confirmed to be a type-II supernova from hydrogen line spectra and whose neutrino output was discovered by four individual detectors \cite{Aglietta, Alekseev, Bionta, Hirata}. Interestingly, the average neutrino energy from SN1987A is lower than contemporary theoretical predictions, and thus the observed neutrino spectrum has a "pinched" maximum not seen in the thermal Fermi-Dirac distribution \cite{Yuksel}. \added{Recalling} the internal energy of the Majorana\added{-Schwinger} gas in the 3D ultra-relativistic case, we find

\begin{align}
\frac{U}{U_0}&\approx 1+2\pi^2 \left(\frac{1}{6}+\frac{1}{\pi^2}(\log 2)^2\right)\frac{T^2}{T_F^2}\notag\\
&=1+2\pi^2 \gamma_3 \frac{T^2}{T_F^2}
\end{align}

\noindent From this ultra-relativistic Majorana correction factor $\gamma_3\approx 0.215$, a low internal energy of the supernova-born neutrino cloud could be a signature that the neutrino \added{gas follows Majorana-Schwinger statistics, and is therefore} a Majorana fermion. Nonetheless, the lack of current supernova neutrino events makes statistical analysis of the thermal distribution difficult; \added{our Majorana-Schwinger} theory predicts a much lower energy than found in the SN1987 spectrum, so it might be possible that a number of these low-energy events originate from the neutrino background \cite{Costantini}. Recent theoretical studies have shown that if neutrinos \added{annihilate then} we would observe faster cooling and lower stellar temperatures, and could better explain the thermal neutrino distribution's "pinched" maximum \cite{Dolgov2,Choubey,Dolgov3}. Because we know that neutrinos have half-integer spin from spin conservation in $\beta$-decay, we could explain this possibly pathological statistical behavior as mutual particle-particle annihilation \added{of Majorana-Schwinger fermions}. Analysis of neutrino emissions from future supernovae could verify if the Majorana\added{-Schwinger} distribution truly describes the statistics of these particles.

\subsubsection{\added{Cosmic Neutrino Background}}
Besides supernovae, one proposed astrophysical source of dense neutrino gases could be the cosmic neutrino background (C$\nu$B), a relic from the early universe when neutrinos decoupled from baryonic matter \cite{Faessler, Zuber}. Currently, the detection of the C$\nu$B are limited to elastic neutrino scattering, neutrino capture by $\beta$-decaying nuclei, and C$\nu $B scattering off cosmic rays \cite{Chiaki}, although direct detection via a large-area surface-deposition tritium source has been proposed \cite{Betts}. If neutrinos are truly Majorana fermions \added{that follow the Majorana-Schwinger statistics}, the temperature of the present-day thermal C$\nu$B would be greatly different than if we considered the neutrino as a "regular" Dirac fermion \cite{Bhattacharjee}. As such, deviations of the C$\nu$B statistics from the Fermi-Dirac system in regions of high density could prove that neutrinos are Majorana\added{-Schwinger} fermions without the need for neutrinoless double-$\beta$ decay, although currently it seems experimentally unlikely to study the thermodynamics of the C$\nu$B.
Perhaps more promising evidence of a cosmological Majorana\added{-Schwinger} gas could be found in observations of the Big Bang nucleosynthesis (BBN), where certain physical observables are highly dependent on the statistics of the decoupled neutrinos in the early universe \cite{Dolgov1}. The abundance of primordial $^4$He as a function of baryon number density tells us that the neutrino distribution in the early universe diverges from both the bosonic and fermionic systems, hinting that neutrinos do not obey pure Fermi-Dirac statistics. A violation of Fermi statistics might also be present in the number of neutrino species at BBN, which could be a further signature of \added{astrophysical Majorana-Schwinger fermions} \cite{Dolgov2}.

\subsubsection{\added{Neutrino sources}}
Apart from the above examples, a possible test of Majorana\added{-Schwinger} thermodynamics in neutrino matter could be found in accelerator-based neutrino sources such as superbeams, which are based on pion decays in the presence of high proton intensity \cite{Kuno}. Similar matter might be created via a "neutrino factory", which utilizes neutrino emission from muon decay. Such Earth-made systems of high-density neutrinos could lead to more experimentally-realizable neutrino thermodynamics (\added{and, possibly, Majorana-Schwinger statistics}) than supernovae emissions, observations of the cosmic neutrino background, or relics of the big bang nucleosynthesis.

\section{VI. Conclusions}

 In the present literature, there is a clear lack of attention towards \added{the development of a non-interacting,} many-body statistics of Majorana particles. Many resources assume that they either observe the traditional Fermi-Dirac statistics (in the case of the fundamental Majorana fermion \added{in the Standard Model}), the "intermediate" statistics of Haldane and Wu (in the case of the Majorana zero mode with Abelian exchange statistics), \added{or the highly topologically-nontrivial non-Abelian anyon statistics}. \added{Moreover, in \added{those} models often considered for many-body Majorana statistics, we are usually restricted to the strong coupling limit, where random fluctuations (in the case of the Sachdev-Ye-Kitaev model) or quantum entanglement (in the case of the Majorana zero mode) restrict us from considering a non-interacting gas of self-conjugate fermions in the topologically trivial regime.}

 Motivated by theoretical and experimental studies of low dimensional topological media and neutralino dark matter, we have explicitly and exhaustively shown that the presence of mutual particle-particle annihilation in a \added{number-conserving} system described by an effective Eddington-Majorana wave equation manifests itself as a completely new theory of quantum statistics distinct from Fermi-Dirac, Bose-Einstein, or the "intermediate" Haldane-Wu statistics of anyons in the lowest Landau level. \added{These particles, which we have called Majorana-Schwinger fermions, obey the spin-statistics relation, in that mutual pairwise annihilation can only occur at some finite temperature, when the effects of Pauli exclusion are sufficiently suppressed.} Through a combinatorial argument, we have found that the Majorana\added{-Schwinger} distribution function exhibits a finite-temperature discontinuity at the chemical potential in the thermodynamic limit, which in turn leads to a deviation from Fermi-Dirac statistics and a residual entropy of $\log 2$ per particle at zero temperature. {The hallmarks of the Majorana\added{-Schwinger} thermodynamics may be easily verified in a modified Kitaev chain in the presence of a localized condensate of boson pairs. The resulting composite Bose-Fermi pairs preserve a global $U(1)$ gauge symmetry and hence preserve total particle number, while the fermionic degrees of freedom remain self-adjoint.} 
 
 \added{Because we can exactly solve for a closed form of the distribution function, we can make accurate thermodynamic predictions, lending our theory to be easily} applied to \added{contemporary} condensed matter systems. \added{Interestingly,} our new statistics agrees with \added{the} finite-temperature Hill thermodynamics of topological systems, experimental signatures of a possible Majorana Fermi surface in the Kondo insulator SmB$_6$, Monte Carlo simulations of the specific heat in a Kitaev honeycomb model, and Raman spectroscopic data in the spin lattice $\alpha$-RuCl$_3$. Our model of the free Majorana\added{-Schwinger} gas also has the potential to yield empirical signals of the Majorana-nature of neutrinos via possible non-Fermi\added{-Dirac} statistics in cosmic neutrino sources.

In terms of future work, the obvious next step is to expand the theory of the Majorana\added{-Schwinger} statistics to interacting many-body ensembles, where the authors expect applications to more realistic systems \cite{Heath}. Similar applications might be found in exotic many-body states in the cosmological limit, where Majorana\added{-Schwinger} statistics becomes a new tool to analyze astrophysical data \cite{Heath}.


\section{Acknowledgements}

\noindent There are countless people who assisted in the development of the Majorana\added{-Schwinger} statistics. One of the authors (J.T.H.) would like to specifically thank Matthew Gochan, whose insights helped form the backbone of the theory, and Emilio Cobanera, who supplied useful information for the present theory's genesis. We would also like to thank Gennady Chitov for stimulating conversation; Jan Engelbrecht, Matthew Heine, Peter Johnson, and Tong Yang for useful theoretical input; Kenneth Burch, Mason Gray, Fazel Tafti, Yiping Wang, and Ilija Zeljkovic for discussions on experimental feasibility; Ron Rubin, Avadh Saxena, and Xu Yang for a thorough review and critique of the paper; and  Stefanos Kourtis, Debanjan Chowdhury, and Suchitra Sebastian for useful commentary and references. Finally, we would like to thank Andr\'e-Marie Tremblay for his useful and crucial comments on the microscopic theory of the Fermi-Bose Hamiltonian described in this paper.  J.T.H. would also like to thank the organizers of the 2018 Gordon Research Conference on Correlated Electron Systems at Mount Holyoke College, where several stimulating conversations at the conference ultimately led to the final version of this paper. This work was partially supported by the John H. Rourke endowment fund at Boston College.


\section{Appendix A. Evaluation of the hypergeometric function's residue for general particle number}

In this Appendix, we explain in more detail the derivation of Eqn. \eqref{13}. We begin by expressing the residue in terms of a geometric series\footnote{A crucial step in this calculation was provided with the help of Greg Martin (\url{https://math.stackexchange.com/users/16078/greg-martin}), from Infinite Sum of Falling Factorial and Power, URL (version: 2016-06-11): \url{https://math.stackexchange.com/q/1821726}; the authors thank \url{math.stackexchange.com} for providing a forum where we could inquire about and conduct research of some of the more mathematical techniques for this and other derivations in our paper}:\\
\begin{align}
&\textrm{Res}_1\left(
\frac{x^{G_j}}{(1-x/2)(1-x)^y}\right)\notag\\
&=\lim_{x\rightarrow 1}\frac{1}{(y-1)!}\frac{d^{y-1}}{dx^{y-1}}\left(\frac{x^{G_j}}{1-x/2}
\right)\notag\\
&=\sum_{k=0}^\infty \frac{1}{2^k}{G_j+k\choose y-1}\label{14}
\end{align}

\noindent By rewriting the sum of binomial coefficients in Eqn. \eqref{14} in terms of a generalized beta function \cite{NIST}, we effectively derive Eqn. \eqref{13}.

{ \section{Appendix B. Anti-commutation and commutation relations of the modified Majorana ladder operators}

Recall the Majorana\added{-Schwinger} operators we defined in Section III C:

\begin{align}
\widetilde{\gamma}_j^\dagger =c_j^\dagger +c_j b_j^\dagger b_j^\dagger ,\qquad \widetilde{\gamma}_j=c_j+c_j^\dagger b_j b_j
\end{align}

To elucidate further the quantum statistics described by such a quantization condition, let us calculate the commutation and anti-commutation relations between these operators. We begin with the latter:

\begin{align}
\{\widetilde{\gamma}_i,\,\widetilde{\gamma}_j^\dagger\}&=\{c_i+c_i^\dagger b_i b_i,\,c_j^\dagger +c_j b_j^\dagger b_j^\dagger \}\notag\\
&=\{c_i,\,c_j^\dagger\}+\{c_i^\dagger b_i b_i,\,c_j b_j^\dagger b_j^\dagger\}\notag\\
&=\delta_{ij}+c_i^\dagger c_j b_i b_i b_j^\dagger b_j^\dagger +c_j c_i^\dagger b_j^\dagger b_j^\dagger b_i b_i \notag\\
&=\delta_{ij} \left\{
1+2c_i^\dagger c_j \left(1+2b_j^\dagger b_i\right)+b_j^\dagger b_j^\dagger b_i b_i
\right\}
\\
\notag\\
\{\widetilde{\gamma}_i,\,\widetilde{\gamma}_j\}&=\{ 
 c_i+c_i^\dagger b_i b_i , \, c_j+c_j^\dagger b_j b_j 
\}
\notag\\
&=\{c_i,\,c_j^\dagger b_j b_j \}+\{c_i^\dagger b_i b_i,\,c_j\}\notag\\
&=\delta_{ij} (b_j b_j+b_i b_i)\\
\notag\\
\{\widetilde{\gamma}_i^\dagger,\,\widetilde{\gamma}_j^\dagger \}&=\{
c_i^\dagger +c_i b_i^\dagger b_i^\dagger,\,c_j^\dagger +c_j b_j^\dagger b_j^\dagger
\}\notag\\
&=\{c_i^\dagger,\,c_j b_j^\dagger b_j^\dagger\}+\{c_i b_i^\dagger b_i^\dagger,\,c_j^\dagger\}\notag\\
&=\delta_{ij}\left(b_j^\dagger b_j^\dagger +b_i^\dagger b_i^\dagger \right)
\end{align}
The first anti-commutation relation is equivalent to the fermionic relation for $i\not=j$. For $i=j$, we see that the first anti-commutation relation yields some finite number dependent on the bosonic and fermionic contributions $n_b$ and $n_f$, respectively:
\begin{align}
\{\widetilde{\gamma}_i^\dagger,\,\widetilde{\gamma}_j\}=1+2n_f(1+2n_b)+n_b(n_b-1)
\end{align}

\begin{table*}
\begin{centering}
\caption{\label{tab:table3}  Comparison of the commutation and anti-commutation relations for the Fermi-Dirac, Bose-Einstein, Kitaev-Majorana, and \added{Majorana-Schwinger} systems. Especially in the anti-commutation relations, we see fermionic-like behavior when $i\not=j$ but bosonic characteristics when the Majorana\added{-Schwinger} fermions occupy the same site.}
\hspace{-10mm}\scalebox{0.85}{\begin{tabular}{l  l l  l }
\hline \hline \\
Fermi-Dirac & Bose-Einstein & Kitaev-Majorana &  "True" Majorana \\
\hline \\
 $\{c_i,\,c_j^\dagger\}=\delta_{ij}\phantom{-2c_j^\dagger c_i} $  
& $\{b_i,\,b_j^\dagger\}=\delta_{ij}+2b_j^\dagger b_i$ 
& $\{\gamma_i,\,\gamma_j^\dagger\}=2\delta_{ij}$
& $\{\widetilde{\gamma}_i,\,\widetilde{\gamma}_j^\dagger\}=\begin{array} {lll} &\delta_{ij}\bigg\{
1+2c_i^\dagger c_j (1+2b_j^\dagger b_i)\\
&\phantom{\delta_{ij}\bigg\{}+b_j^\dagger b_j^\dagger b_i b_i 
\bigg\}\end{array}$
\\\\
$\{c_i,\,c_j\}=0$ & $\{b_i,\,b_j\}=2b_ib_j$ & $\{\gamma_i,\,\gamma_j\}=2\delta_{ij}$ & $\{\widetilde{\gamma}_i,\,\widetilde{\gamma}_j\}=\delta_{ij}(b_jb_j+b_ib_i)$ \\\\
$\{c_i^\dagger ,\,c_j^\dagger\}=0$ & $\{b_i^\dagger,\,b_j^\dagger\}=2b_i^\dagger b_j^\dagger$ & $\{\gamma_i^\dagger,\,\gamma_j^\dagger\}=2\delta_{ij}$ & $\{\widetilde{\gamma}_i^\dagger,\,\widetilde{\gamma}_j^\dagger\}=\delta_{ij}(b_j^\dagger b_j^\dagger +b_i^\dagger b_i^\dagger)$\\\\
\hline\hline  \\
$[c_i,\,c_j^\dagger]=\delta_{ij}-2c_j^\dagger c_i $& $[b_i,\,b_j^\dagger]=\delta_{ij}$ & 
 $[\gamma_i,\,\gamma_j^\dagger ]=  \begin{array} {lll} & 2(c_i^\dagger c_j -c_j^\dagger c_i) \\ & +2(c_ic_j+c_i^\dagger c_j^\dagger) \end{array}  $
  & 
		    $[\widetilde{\gamma}_i,\,\widetilde{\gamma}_j^\dagger ]=\begin{array} {lll} & \delta_{ij}+2(c_i^\dagger c_j\delta_{ij}-c_j^\dagger c_i) \\  & +2(c_i c_j b_j^\dagger b_j^\dagger +c_i^\dagger c_j^\dagger b_i b_i) \\  &+\delta_{ij}(4c_i^\dagger c_jb_j^\dagger b_i-b_j^\dagger b_j^\dagger b_i b_i)\\ & +2c_i^\dagger c_j b_j^\dagger b_j^\dagger b_i b_i \end{array}$ 
 \\\\
 		    $[c_i,\,c_j]=2c_i c_j\phantom{-\delta_{ij}^\dagger\,}$ & $[b_i,\,b_j]=0$ & $[\gamma_i,\,\gamma_j]=\begin{array} {lll} & 2(c_i^\dagger c_j -c_j^\dagger c_i) \\ & +2(c_ic_j+c_i^\dagger c_j^\dagger) \end{array} $ 
		    &
		     $[\widetilde{\gamma}_i,\,\widetilde{\gamma}_j]
		    =\begin{array}{lll} &2c_i c_j +\delta_{ij}(b_j b_j -b_i b_i)\\ &+2(c_i^\dagger c_j b_i b_i -c_j^\dagger c_i b_j b_j) \\ &+2c_i^\dagger c_j^\dagger b_i b_i b_j b_j \end{array}$\\\\
		    $[c_i^\dagger,\,c_j^\dagger]=2c_i^\dagger c_j^\dagger\phantom{-\delta_{ij}\,}$ & $[b_i^\dagger,\,b_j^\dagger]=0$ &  $[\gamma_i^\dagger,\,\gamma_j^\dagger]=\begin{array} {lll} & 2(c_i^\dagger c_j -c_j^\dagger c_i) \\ & +2(c_ic_j+c_i^\dagger c_j^\dagger) \end{array} $ 
		    &
		     $[\widetilde{\gamma}_i,\,\widetilde{\gamma}_j]
		    =\begin{array}{lll} &2c_i^\dagger c_j^\dagger +\delta_{ij} (b_i^\dagger b_i^\dagger -b_j^\dagger b_j^\dagger)\\  &+2(c_i^\dagger c_j b_j^\dagger b_j^\dagger -c_j^\dagger c_i b_i^\dagger b_i^\dagger) \\ &+2c_i c_j b_i^\dagger b_i^\dagger b_j^\dagger b_j^\dagger\end{array}$  \\\\ \hline \hline
\end{tabular}}
\end{centering}
\end{table*}

This is similar to the bosonic case, in that the anti-commutation relation is dependent upon the population of bosons on the site in question. More pronounced bosonic-like behavior is seen in the last two anti-commutation relations, which yield zero if $i\not=j$ (as in the traditional fermionic system), but are equivalent to the bosonic anti-commutation relation for $i=j$. Indeed, in the latter case, we see $\{\widetilde{\gamma}_i,\,\widetilde{\gamma}_i\}=2\widetilde{\gamma}_i\widetilde{\gamma}_i=2b_i b_i$, implying that the \added{Majorana-Schwinger} destruction operator has the same effect as a bosonic destruction operator on the Hilbert space of the system if the site we act the operator on is already occupied. A similar relationship is seen for the \added{Majorana-Schwinger} creation operators. In this way, the \added{Majorana-Schwinger} operators are seen to be similar to the operators of a Kitaev-Majorana zero mode. From the self-adjoint nature of the Kitaev-Majorana operator, the anti-commutation relation between creation or annihilation operators for $i=j$ will yield a non-zero result, in contrast to the fermionic case. In our system we also have a non-zero result for any anti-commutation relation when $i=j$, but we see that such a scenario is dependent on the bosonic degrees of freedom present in the system. A comparison of the anti-commutation relations between the Fermi-Dirac, Bose-Einstein, Kitaev-Majorana, and \added{Majorana-Schwinger} second quantization operators is shown in Table \ref{tab:table3}. 
%
%
We now proceed to calculate the commutation relations of the \added{Majorana-Schwinger} operators, which are somewhat more non-trivial then the anti-commutation relations:

\begin{align}
[\widetilde{\gamma}_i,\,\widetilde{\gamma}_j^\dagger]&=[c_i+c_i^\dagger b_i b_i,\,c_j^\dagger +c_j b_j^\dagger b_j^\dagger]\notag\\
&=[c_i,\,c_j^\dagger]+[c_i,\,c_jb_j^\dagger b_j^\dagger]+[c_i^\dagger b_i b_i,\,c_j^\dagger]+[c_i^\dagger b_i b_i,\,c_j b_j^\dagger b_j^\dagger]\notag\\
&=\delta_{ij}-2c_j^\dagger c_i +2c_i c_j b_j^\dagger b_j^\dagger +2c_i^\dagger c_j^\dagger b_i b_i+c_i^\dagger c_j b_i b_i b_j^\dagger b_j^\dagger-c_j c_i^\dagger b_j^\dagger b_j^\dagger b_i b_i\notag\\
&=\delta_{ij}+2(c_i^\dagger c_j \delta_{ij}-c_j^\dagger c_i)+2(c_i c_j b_j^\dagger b_j^\dagger +c_i^\dagger c_j^\dagger b_i b_i)+\delta_{ij}(4b_j^\dagger b_i c_i^\dagger c_j-b_j^\dagger b_j^\dagger b_i b_i)+c_i^\dagger c_j b_j^\dagger b_j^\dagger b_i b_i
\end{align}
\begin{align}
[\widetilde{\gamma}_i,\,\widetilde{\gamma}_j]&=[c_i+c_i^\dagger b_i b_i,\,c_j+c_j^\dagger b_j b_j]\notag\\
&=[c_i,\,c_j]+[c_i,\,c_j^\dagger b_j b_j]+[c_i^\dagger b_i b_i,\,c_j]+[c_i^\dagger b_i b_i ,\,c_j^\dagger b_j b_j]\notag\\
&=2c_i c_j+(\delta_{ij}-2c_j^\dagger c_i)b_j b_j +(2c_i^\dagger c_j-\delta_{ij})b_i b_i +2c_i^\dagger c_j^\dagger b_i b_i b_j b_j\notag\\
&=2c_i c_j +\delta_{ij}(b_j b_j -b_i b_i)+2(c_i^\dagger c_j b_i b_i -c_j^\dagger c_i b_j b_j)+2c_i^\dagger c_j^\dagger b_i b_i b_j b_j
\end{align}
\begin{align}
[\widetilde{\gamma}_i^\dagger,\,\widetilde{\gamma}_j^\dagger]&=[c_i^\dagger+c_i b_i^\dagger b_i^\dagger,\,c_j^\dagger+c_j b_j^\dagger b_j^\dagger]\notag\\
&=[c_i^\dagger,\,c_j^\dagger]+[c_i^\dagger,\,c_j b_j^\dagger b_j^\dagger]+[c_i b_i^\dagger b_i^\dagger,\,c_j^\dagger]+[c_i b_i^\dagger b_i^\dagger ,\,c_j b_j^\dagger b_j^\dagger]\notag\\
&=2c_i^\dagger c_j^\dagger +(2c_i^\dagger c_j -\delta_{ij})b_j^\dagger b_j^\dagger +(\delta_{ij}-2c_j^\dagger c_i)b_i^\dagger b_i^\dagger+2c_i c_j b_i^\dagger b_i^\dagger b_j^\dagger b_j^\dagger \notag\\
&=2c_i^\dagger c_j^\dagger +\delta_{ij} (b_i^\dagger b_i^\dagger -b_j^\dagger b_j^\dagger) +2(c_i^\dagger c_j b_j^\dagger b_j^\dagger -c_j^\dagger c_i b_i^\dagger b_i^\dagger)+2c_i c_j b_i^\dagger b_i^\dagger b_j^\dagger b_j^\dagger
\end{align}

\noindent Note that the Kitaev-Majorana commutation relation $[\gamma_i,\,\gamma_j]$ is zero for $i=j$, while the \added{Majorana-Schwinger} commutation relation $[\widetilde{\gamma}_i,\,\widetilde{\gamma}_j^\dagger]$ is non-zero for $i=j$. The other two commutation relations for the \added{Majorana-Schwinger} operators are zero for $i=j$ and non-zero for $i\not=j$, much like the original Kitaev formulation. A comparison of the different commutation relations is shown in Table \ref{tab:table3}.}

\section{Appendix C. The Incomplete Fermi-Dirac Function}

We want to find a simple form for the incomplete Fermi-Dirac function:

\begin{align}
    F_{\gamma+1}(\mu/T,\,\mu)=\frac{1}{\Gamma(\gamma+1)}\int_{\mu}^\infty \frac{\epsilon^\gamma}{e^{(\epsilon-\mu)/T}+1}d\epsilon
\end{align}\\
\vspace{1mm}

This integral is exceedingly difficult to analyze, so let's look at the low-temperature limit. We make the substitution $xT=\epsilon-\mu$, which transforms the above into 
\begin{align}
F_{\gamma+1}\left(\mu/T,\,\mu\right)&=\frac{T\mu^\gamma}{\Gamma(\gamma+1)}\int_0^\infty \frac{(xT/\mu+1)^\gamma}{e^x+1}dx
\end{align}
We can now utilize the form of the binomial expansion and the Gamma function to simplify a portion of the above \cite{NIST}:
\begin{align}
&\frac{1}{\Gamma(\gamma+1)}\left(xT/\mu+1\right)^\gamma\notag\\
&=\frac{1}{\Gamma(\gamma+1)}+\sum_{k=1}^\infty (-1)^k \frac{(-\gamma)_k}{k\gamma!}\left(\frac{xT}{\mu}\right)^k\frac{1}{\Gamma(k)}\notag\\
&=\frac{1}{\Gamma(\gamma+1)}+\sum_{k=1}^\infty \frac{1}{k(\gamma-k)!}\left(\frac{xT}{\mu}\right)^k
\end{align}
Which, in turn, tells us that we can express the incomplete Fermi-Dirac function as an infinite sum of complete Fermi-Dirac functions:
\begin{align}
F_{\gamma+1}(\mu/T,\,\mu)&=\int_0^\infty \frac{T\mu^\gamma}{e^x+1}\frac{1}{\Gamma(\gamma+1)}\left(xT/\mu+1\right)^\gamma dx\notag\\
&=\int_0^\infty \frac{T\mu^\gamma}{e^x+1}\left(
\frac{1}{\Gamma(\gamma+1)}+\sum_{k=1}^\infty \frac{1}{k(\gamma-k)!}\left(\frac{xT}{\mu}\right)^k\frac{1}{\Gamma(k)}
\right) dx\notag\\
&=T\mu^\gamma \frac{\log 2}{\Gamma(\gamma+1)}+\sum_{k=1}^\infty \frac{1}{(\gamma-k)!}T^{k+1}\mu^{\gamma-k}F_{k+1}(0,\,0)\notag\\
&=\sum_{k=0}^\infty \frac{1}{(\gamma-k)!}T^{k+1}\mu^{\gamma-k}F_{k+1}(0,\,0)
\end{align}

\section{Appendix D. Derivation of the thermodynamic observables in the Majorana\added{-Schwinger} gas}

Recall the approximate relation found between the Fermi energy and chemical potential in the text:

\begin{align}
\frac{2}{3}\epsilon_F^{3/2}\approx\frac{2}{3}\mu^{3/2}+T\mu^{1/2} \log 2 +\frac{\pi^2}{24}\frac{T^2}{\mu^{1/2}}
\end{align}

\noindent We can solve for the chemical potential by suggesting a form of $\mu=\epsilon_F(1+\delta\mu)$:

\begin{align}
1&\approx \frac{\mu^{3/2}}{\epsilon_F^{3/2}}+T\frac{\mu^{1/2}}{\epsilon_F^{3/2}}\frac{3}{2}\log 2+\frac{\pi^2}{16}\frac{T^2}{\epsilon_F^{3/2}\mu^{1/2}}\notag\\
&\approx (1+\delta \mu)^{3/2}+(1+\delta\mu)^{1/2}\frac{3}{2}\frac{T}{T_F} \log 2 +\frac{\pi^2}{16}\frac{T^2}{T_F^{2}}\frac{1}{(1+\delta \mu)^{1/2}}\notag\\
&\approx \left(1+\frac{3}{2}\delta \mu\right)+\frac{3T}{2T_F}\log 2 \left(1+\frac{1}{2}\delta \mu\right)+\frac{\pi^2}{16}\frac{T^2}{T_F^2}\left(1-\frac{1}{2}\delta \mu\right)
\end{align}
Rearranging the above, we find that $\delta \mu$ is given by

\begin{align}
\delta \mu&=\frac{
-\frac{3T}{2T_F}\log 2 -\frac{\pi^2}{16}\frac{T^2}{T_F^2}
}{
\frac{3}{2}+\frac{3T}{4T_F}\log 2-\frac{\pi^2}{32}\frac{T^2}{T_F^2}
}\notag\\
&\approx-\frac{T}{T_F}\log 2 -\left(\frac{\pi^2}{24}-\frac{(\log 2)^2}{2}\right)\frac{T^2}{T_F^2}
\end{align}
Hence, up to second order in temperature, the chemical potential for the non-interacting Majorana\added{-Schwinger gas} is given approximately by

\begin{align}
\mu\approx \epsilon_F \left(
1-\log 2 \frac{T}{T_F}-\left(
\frac{\pi^2}{24}-\frac{(\log 2)^2}{2}
\right)\frac{T^2}{T_F^2}
\right)\label{21}
\end{align}
We thus see that the Majorana\added{-Schwinger} chemical potential is characterized by a linear temperature dependence unseen in the fermionic system. 

We proceed in the same fashion for the finite-temperature internal energy $U(T)$ of the Majorana\added{-Schwinger} system. In terms of the incomplete Fermi-Dirac function, 

\begin{align}
U&\sim V \left(
\frac{2}{5}\mu^{5/2}
+\Gamma(5/2)F_{5/2}(\mu/T,\,\mu)
\right)
\end{align}

\noindent From Eqn. \eqref{21}, we see that

\begin{align}
F_{5/2}(\mu/T,\,\mu)\approx T \mu^{3/2} \frac{\log 2}{\Gamma(5/2)}+\frac{\pi^{3/2}}{6}T^2 \mu^{1/2}
\end{align}

We can now calculate the energy density with the help of the chemical potential in Eqn. \eqref{21}:

\begin{align}
u&\sim \left(
\frac{2}{5}\mu^{5/2}+T\mu^{3/2}\log 2+T^2 \mu^{1/2}\frac{\pi^2}{8}
\right)\notag\\
&= \bigg\{
\frac{2}{5}\epsilon_F^{5/2}\left(
1-\frac{T}{T_F}\log 2 -\left(
\frac{\pi^2}{24}-\frac{(\log 2)^2}{2}
\right)\frac{T^2}{T_F^2}
\right)^{5/2}\notag\\
&\phantom{}+T\epsilon_F^{3/2}\log 2\left(
1-\frac{T}{T_F}\log 2 -\left(
\frac{\pi^2}{24}-\frac{(\log 2)^2}{2}
\right)\frac{T^2}{T_F^2}
\right)\notag\\
&\phantom{}+\frac{\pi^2}{8}T^2 \epsilon_F^{1/2}
\left(
1-\frac{T}{T_F}\log 2 -\left(
\frac{\pi^2}{24}-\frac{(\log 2)^2}{2}
\right)\frac{T^2}{T_F^2}
\right)
\bigg\}\notag\\
&\approx\frac{3}{5}n\epsilon_F \left\{1+\left(\frac{5\pi^2}{24}-\frac{5}{8}(\log 2)^2\right)\frac{T^2}{T_F^2}\right\}
\end{align}

\noindent Interestingly, although the Majorana\added{-Schwinger} gas' chemical potential differs greatly from that of the Fermi\added{-Dirac} gas, the Majorana\added{-Schwinger} energy density follows a fermionic temperature dependence. We can see this be introducing the three-dimensional correction term

\begin{align}
\gamma_3 =\frac{1}{2}-\frac{3}{2\pi^2}(\log 2)^2\label{22}
\end{align}
Now, the energy density is identical to its fermionic counterpart, except now with the $T^2$ term reduced by a factor of $\gamma$:
\begin{align}
u&=\frac{3}{5}n\epsilon_F \left(1+\frac{5\pi^2}{12}\gamma_3 \frac{T^2}{T_F^2}\right)
\end{align}
As a consequence, the specific heat and pressure also behave in a fermionic fashion:
\begin{align}
C_v&=\frac{\pi^2}{2}n\gamma_3  \frac{T}{T_F}
\end{align}
\begin{align}
P&=\frac{2}{3}u=\frac{2}{5}n\epsilon_F \left(1+\frac{5\pi^2}{12}\gamma_3 \frac{T^2}{T_F^2}\right)
\end{align}
The entropy density might be found via the fundamental thermodynamic relations:
\begin{align}
s&=\frac{u+P-n\mu}{T}\notag\\
&\approx \bigg\{
\frac{\pi^2n}{4}\left(\frac{1}{2}-\frac{3}{2\pi^2}(\log 2)^2\right)+\frac{\pi^2 n}{6}\left(\frac{1}{2}-\frac{3}{2\pi^2}(\log 2)^2\right)\notag\\
&\phantom{=\bigg(}+n\left(\frac{\pi^2}{24}-\frac{(\log 2)^2}{2}\right)
\bigg\}\frac{T}{T_F}+n\log 2 \notag\\
&=n\left(
\left(\frac{\pi^2}{4}-\frac{9}{8}(\log 2)^2\right)\frac{T}{T_F}+\log 2
\right)
\end{align}
We therefore find a simple solution to the entropy per particle in terms of the Majorana\added{-Schwinger} correction term Eqn. \eqref{22}:
\begin{align}
\frac{S}{N}=\left(\frac{3\pi^2}{4}\gamma_3-\frac{\pi^2}{8}\right)\frac{T}{T_F}+\log 2
\end{align}
It is important to note here that the above is only true if we consider small, finite temperature. This term is interesting, because it implies that the entropy is non-zero for a zero-temperature system. This residual entropy term is discussed in detail in the main text of the paper.

We continue into the two-dimensional Majorana\added{-Schwinger} system. As in the three-dimensional Majorana\added{-Schwinger} system, the zero-temperature chemical potential is identical to that of the \added{Fermi-Dirac} system. For finite temperatures, we utilize the incomplete Fermi-Dirac function to find a simple equation connecting the chemical potential with the Fermi energy:
\begin{align}
\epsilon_F=\mu+F_1(\mu/T,\,\mu)
\end{align}
This function is easily found with Eqn. \eqref{21}:
\begin{align}
F_1(\mu/T,\,\mu)&=\sum_{k=0}^\infty \frac{1}{(-k)!}T^{k+1}\mu^{-k}F_{k+1}(1)\notag\\
&=T\log 2
\end{align}
The 2D Majorana\added{-Schwinger} chemical potential is thus trivially found:
\begin{align}
\mu&=\epsilon_F \left(1-\frac{T}{T_F}\log 2\right)\label{23}
\end{align}
Much like the two-dimensional Fermi\added{-Dirac} gas, we have a closed, exact form for the two-dimensional chemical potential. 

The energy follows similarly from our form of the incomplete Fermi-Dirac function:
\begin{align}
u&=\int_0^\infty \epsilon g(\epsilon) n_{p\sigma} d\epsilon\notag\\
&=\frac{1}{2}n\epsilon_F\left(\frac{1}{2}\mu^2 +F_2(\mu/T,\mu)\right)\notag\\
&=\frac{1}{4}n\epsilon_F\left(
\mu^2 +2\mu T \log 2 +\frac{\pi^2}{6}T^2
\right)
\end{align}

We now plug in the chemical potential Eqn. \eqref{23} to obtain a closed form for the 2D Majorana\added{-Schwinger internal} energy:

\begin{align}
u&=\frac{1}{2}n\epsilon_F \left\{
1+\left(
\frac{\pi^2}{6}-(\log 2)^2
\right)\frac{T^2}{T_F^2}
\right\}\label{24}
\end{align}
We can simplify this by introducing the two-dimensional \added{Majorana-Schwinger} correction term

\begin{align}
\gamma_2=\frac{1}{2}-\frac{3}{\pi^2}(\log 2)^2
\end{align}
Thus, Eqn. \eqref{24} becomes

\begin{align}
u=\frac{1}{2}n\epsilon_F \left(1+\frac{\pi^2}{3}\gamma_2 \frac{T^2}{T_F^2}\right)
\end{align}
From the above, the specific heat and pressure follow:

\begin{align}
C_v=\frac{\pi^2}{3}n \gamma_2 \frac{T}{T_F}
\end{align}

\begin{align}
P=\frac{1}{2}n\epsilon_F \left(
1+\frac{\pi^2}{3}\gamma_2 \frac{T^2}{T_F^2}
\right)
\end{align}
The entropy per particle is then obtained with the same method as before:
\begin{align}
\frac{S}{N}&=\frac{\pi^2}{3}\gamma_2 \frac{T}{T_F}+\log 2
\end{align}
In addition to the 3D and 2D cases, we calculate the thermodynamic observables of the 1D Majorana\added{-Schwinger} gas. For the sake of brevity, we simply quote the results:

\begin{subequations}
\begin{equation}
\mu\approx \epsilon_F \left(
1-\frac{T}{T_F}\log 2 +\left(\frac{\pi^2}{24}-\frac{(\log 2)^2}{2}\right)\frac{T^2}{T_F^2}
\right)
\end{equation}\\
\begin{equation}
u\approx \frac{1}{3}n\epsilon_F \left(1+\frac{\pi^2}{4}\gamma_1\right)
\end{equation}
\begin{equation}
C_v \approx \frac{\pi^2}{6}N \gamma_1 \frac{T}{T_F}
\end{equation}
\begin{equation}
S\approx \frac{\pi^2}{6}\gamma_1 \frac{T}{T_F}+\log 2
\end{equation}
\end{subequations}
Where the 1D Majorana\added{-Schwinger} correction term is given by

\begin{align}
\gamma_1=\frac{1}{2}-\frac{6}{\pi^2}(\log 2)^2
\end{align}

\noindent The calculation of the above quantities is nearly identical to the 3D case.\\ \vspace{2mm}
\newpage 
\vspace{7mm}


\bibliographystyle{iopart-num}
\bibliography{main}{}

\end{document}